%% file: main.tex
\RequirePackage[prologue,cmyk]{xcolor}
\documentclass[acmsmall,nonacm,screen,prologue,svgnames,acmthm]{acmart}
\pdfoutput=1

\usepackage[noscreen]{clrscode4e}
\usepackage{mathtools}
\usepackage{bbm}
\usepackage{listings}
\usepackage{subcaption}
\usepackage[inline]{enumitem}
\usepackage[capitalize]{cleveref}
\usepackage{booktabs}
\usepackage{wrapfig}
\usepackage{multirow}
\usepackage{quantikz}

\usepackage{pifont}

\input{prelude}
\input{styles}

\begin{document}

\title{Modular Synthesis of Efficient Quantum Uncomputation}

\input{authorinfo}
\input{abstract}

\maketitle

\input{introduction}
\input{background}
\input{language}
\input{implementation}

\input{evaluation}
\input{conclusion}

\input{acknowledgment}

\newpage
\bibliographystyle{ACM-Reference-Format}
\bibliography{main}

\newpage
\appendix
\section{Proof of Uncomputation Synthesis}
\label{appendix:proofs-uncomp}
\input{proofs-uncomp}

\end{document}

%% file: prelude.tex
\DeclareMathOperator{\opIf}{\mathbf{if}}

\newcommand\set[1]{\ensuremath{\left\{#1\right\}}}

\newcommand\hilbert[1]{\ensuremath{\mathcal{H}\left(#1\right)}}

\newcommand\abs[1]{\ensuremath{\left\lvert #1 \right\rvert}}

\DeclareMathOperator\plusplus{+\hspace{-0.3em}+}

\newcommand*{\FVar}[1]{\texttt{{#1}}}
\newcommand*{\CVar}[1]{\mathit{{#1}}_\texttt{c}}
\newcommand*{\QVar}[1]{\mathit{{#1}}}

\newcommand*{\SelOp}{\gamma}
\newcommand*{\UnSelOp}{\rotatebox[origin=c]{180}{$\gamma$}}
\newcommand*{\NewOp}[1]{\FVar{new}_{#1}}
\newcommand*{\UnNewOp}[1]{\FVar{del}_{#1}}
\newcommand*{\CatOp}[1]{\FVar{concat}_{#1}}
\newcommand*{\UnCatOp}[1]{\CatOp{#1}^\dagger}
\newcommand*{\DupOp}{\FVar{dup}}
\newcommand*{\UnDupOp}{\FVar{undup}}
\newcommand*{\PhaseOp}[1]{{\FVar{phase}_{#1}}}
\newcommand*{\ForgetOp}{\FVar{forget}}
\newcommand*{\MeasureOp}{\FVar{measure}}
\newcommand*{\HadamardOp}{\FVar{H}}
\newcommand*{\DisposeOp}{\FVar{dispose}}

\newcommand*{\Stmt}[1]{\big\langle #1 \big\rangle}

%% file: styles.tex
\AtEndPreamble{%
  \theoremstyle{acmdefinition}
  \newtheorem{property}[theorem]{Property}%
  \crefname{property}{Property}{Properties}%
}

%% file: authorinfo.tex
\author{Hristo Venev}
\affiliation{%
  \institution{INSAIT, Sofia University ``St. Kliment Ohridski''}
  \city{Sofia}
  \country{Bulgaria}}
\email{hristo.venev@insait.ai}

\author{Timon Gehr}
\affiliation{%
  \institution{ETH Zurich}
  \city{Zurich}
  \country{Switzerland}}
\email{timon.gehr@inf.ethz.ch}

\author{Dimitar Dimitrov}
\affiliation{%
  \institution{INSAIT, Sofia University ``St. Kliment Ohridski''}
  \city{Sofia}
  \country{Bulgaria}}
\email{dimitar.dimitrov@insait.ai}

\author{Martin Vechev}
\affiliation{%
  \institution{ETH Zurich}
  \city{Zurich}
  \country{Switzerland}}
\email{martin.vechev@inf.ethz.ch}

%% file: abstract.tex
\begin{abstract}
A key challenge of quantum programming is uncomputation: the reversible deallocation of qubits. And while there has been much recent progress on automating uncomputation, state-of-the-art methods are insufficient for handling today’s expressive quantum programming languages. A core reason is that they operate on primitive quantum circuits, while quantum programs express computations beyond circuits, for instance, they can capture families of circuits defined recursively in terms of uncomputation and adjoints.

In this paper, we introduce the first modular automatic approach to synthesize correct and efficient uncomputation for expressive quantum programs. Our method is based on two core technical contributions: (i) an intermediate representation (IR) that can capture expressive quantum programs and comes with support for uncomputation, and (ii) modular algorithms over that IR for synthesizing uncomputation and adjoints.

We have built a complete end-to-end implementation of our method, including an implementation of the IR and the synthesis algorithms, as well as a translation from an expressive fragment of the Silq programming language to our IR and circuit generation from the IR. Our experimental evaluation demonstrates that we can handle programs beyond the capabilities of existing uncomputation approaches, while being competitive on the benchmarks they can handle. More broadly, we show that it is possible to benefit from the greater expressivity and safety offered by high-level quantum languages without sacrificing efficiency.
\end{abstract}

%% file: introduction.tex
\section{Introduction} \label{sec:intro}
Classical programs often change data irreversibly.
For example, when a variable is assigned a new value,
then
the old one is erased from memory
and it cannot be recovered anymore.
In contrast,
quantum programs can rarely afford to destroy information
and primarily use reversible operations.
Assignment is commonly emulated with the controlled \proc{not} gate $y \gets x \oplus y$,
which sets $y$ to the exclusive or of the two variables.
If $y$ is guaranteed to equal zero beforehand,
then the gate effectively acts as an assignment.
However, ensuring reversibility complicates many otherwise simple tasks.
Notably, temporary variables have to be deallocated reversibly
in order to preserve quantum interference patterns \citep{shor}.
This process is known as \emph{uncomputation},
and involves resetting a variable to its \emph{zero state}
by undoing all changes to it, last to first \citep{lecerf_machines_1963,bennett_logical_1973,toffoli_reversible_1980}.
Once reset, the variable can be marked as free for later use.
While uncomputation may sound straightforward,
programming it manually is tedious and error-prone,
which makes it a major practical issue,
and therefore, it is crucial to automate it as much as possible.

\begin{figure}
  \begin{subfigure}{0.45\textwidth}
    \begin{codebox}
      \Procname{$\proc{Iterate}(n, f, \kw{in}\ x, \kw{out}\ y)$} \li
      \If $n = 0$:\, $y \gets x \oplus y$ \li
      \Then
      \Else
        \kw{alloc} $t$ \li
        $\proc{Iterate}(n-1, f, x, t)$ \li
        $f(t, y)$ \li
        $\proc{Iterate$^\dag$}(n-1, f, x, t)$ \li
        \kw{alloc}$^\dag$ $t$
      \End
    \end{codebox}
    \caption{
      For $n \ge 1$ \proc{Iterate} first computes iteration $n - 1$,
      then applies $f$,
      and finally uncomputes iteration $n - 1$ by calling the adjoint procedure \proc{Iterate$^\dag$}.
    }
    \label{example:naive-uncomputation-iterate}
  \end{subfigure}
  \hfill
  \begin{subfigure}{0.45\textwidth}
    \begin{codebox}
      \Procname{$\proc{Iterate$^{\dag}$}(n, f, \kw{in}\ x, \kw{out}^\dag\ y)$} \li
      \If $n = 0$:\, $y \gets x \oplus y$ \li
      \Then
      \Else
        \kw{alloc} $t$ \li
        $\proc{Iterate}(n-1, f, x, t)$ \li
        $f^\dag(t, y)$ \li
        $\proc{Iterate$^{\dag}$}(n-1, f, x, t)$ \li
        \kw{alloc}$^\dag$ $t$
      \End
    \end{codebox}
    \caption{
      \proc{Iterate}$^\dag$ is almost the same,
      except that it is called with $y$ already equal to the $n$th iteration
      and uncomputes it with $f^\dag(\kw{in}\ x, \kw{out}^\dag\ y)$.
    }
    \label{fig:naive-uncomputation-iterate-adj}
  \end{subfigure}
  \caption{
    \proc{Iterate} reversibly computes the $n$th iteration of the procedure $f(\kw{in}\ x, \kw{out}\ y)$,
    which reads $x$, and updates $y$ in-place.
    Here $n$ and $f$ hold classical data,
    while $x$ and $y$ hold quantum data.
    Input quantum parameters are not modified,
    output ones are expected in a zero state,
    and adjoint output ones are uncomputed to a zero state.
    Allocation and deallocation happens with two primitives:
    \kw{alloc} pops a zeroed quantum register from the free list,
    while its adjoint \kw{alloc$^\dag$} pushes a zeroed quantum register back.
  }
  \label{fig:naive-uncomputation}
\end{figure}

\subsection{Uncomputation is more than adjoints}
The first step towards automating uncomputation is the ability to synthesize \emph{adjoints} $f^\dag$:
the inverses of whole units of code $f$.
Circuit description languages like Quipper \citep{quipper} and Qwire \citep{qwire},
and libraries like Qiskit \citep{qiskit}
can generate adjoints of circuits,
and the Q\# language \citep{qsharp} supports adjoints of recursive procedures.
Although the specifics vary,
the fundamental idea of adjoint synthesis is to rewrite code using the equivalence $(A ; B)^\dag = B^\dag ; A^\dag$
known from linear algebra.
\Cref{fig:naive-uncomputation} illustrates this transformation for the \proc{Iterate} procedure,
which iterates $n$ times another procedure $f$.

Although key,
adjoint synthesis alone does not guarantee correct and efficient uncomputation.

\paragraph{Lack of correctness guarantees}
Extra care is needed to ensure that an adjoint indeed uncomputes a variable.
For example, in the typical compute-use-uncompute pattern $f ; g ; f^\dag$,
if $g$ accidentally modifies a variable computed by $f$,
then the adjoint $f^\dag$ will not undo that update,
and the variable will not be zeroed.
More importantly, even if no updates are missed,
uncomputation can still fail
when the quantum effects of $f$ and $g$ are incompatible (see, e.g., \citet{silq}).
This aspect is overlooked in languages with built-in support for the compute-use-uncompute pattern,
that is, for \emph{conjugation}:
Quipper, Q\#, Tower \cite{tower}, and Qunity \cite{qunity}
all support conjugation but provide no correctness guarantees for uncomputation.

\paragraph{Lack of efficiency guarantees}
Naively implemented uncomputation,
even when correct,
can greatly increase resource usage
and cause an exponential blowup in recursive procedures.
This is the case for \proc{Iterate} in \cref{fig:naive-uncomputation},
which makes $2^{n-1}$ calls to the procedure $f$
and $2^{n-1} - 1$ calls to $f^\dag$,
provided $n \ge 1$.
This happens because to uncompute $f^{n-1}(x)$,
\proc{Iterate} invokes \proc{Iterate$^\dag$},
which in turn recurses again to recompute $f^{n - 2}(x)$.
Such recomputation can be avoided
by switching the base algorithm to a tail recursive scheme,
as noted by \citet{tower}.
However, this transformation can significantly complicate the code and obscure its correctness.

A key challenge then is how to automatically synthesize both correct and efficient uncomputation.

\subsection{Existing work towards uncomputation for quantum programs}
Synthesis of correct and efficient uncomputation was initially developed
to transform irreversible
classical machines \citep{lecerf_machines_1963,bennett_logical_1973}
and circuits \citep{toffoli_reversible_1980}
into reversible ones.
These methods were later utilized for compiling classical programs to quantum circuits
\citep{quipper,parent2015reversible,amy2017verified,reqwire}.
However, more research was needed to extend them to more general quantum programs.

\begin{figure}
  \begin{subfigure}{0.45\textwidth}
    \begin{codebox}
      \Procname{$\proc{Iterate}(n, f, \kw{in}\ x, \kw{out}\ y)$} \li
      \If $n = 0$:\, $y \gets x \oplus y$ \li
      \Then
      \Else
        \kw{alloc} $t$ \li
        $\proc{Iterate}(n-1, f, x, t)$ \li
        $f(t, y)$ \li
        \kw{forget} $t$
      \End
    \end{codebox}
    \caption{A declarative version of \proc{Iterate}.}
    \label{fig:forget-statement-iterate}
  \end{subfigure}
  \hfill
  \begin{subfigure}{0.45\textwidth}
    \begin{codebox}
      \Procname{$\proc{Iterate$^{\textsf{G}}$}(n, f, \kw{in}\ x, \kw{out}\ y, \kw{bin}\ g)$} \li
      \If $n = 0$:\, $y \gets x \oplus y$ \li
      \Then
      \Else
        \kw{alloc} $t$ \li
        $\proc{Iterate$^{\textsf{G}}$}(n-1, f, x, t, g)$ \li
        $f^\textsf{G}(t, y, g)$ \li
        \kw{dispose} $t$ \kw{in} $g$
      \End
    \end{codebox}
    \caption{A garbage-producing version of \proc{Iterate}.}
    \label{fig:forget-statement-iterate-garb}
  \end{subfigure}
  \caption{A \kw{forget} statement can be thought of a placeholder for yet to be synthesized uncomputation.  That uncomputation can be delayed by throwing the data into a garbage bin using \kw{dispose}.}
  \label{fig:forget-statement}
\end{figure}

\subsubsection{Specifying uncomputation}
For more general quantum programs
we need a linguistic device
for specifying the variables to be uncomputed.
The mechanism should have a precise semantics
that abstracts the concrete operational details,
allowing a variety of implementations.
It should also have a clear criterion when
a variable can be uncomputed correctly.
These questions were first investigated in depth
in the high-level quantum programming language Silq \citep{silq}.
The key idea of Silq is to separate the specification of uncomputation from its implementation.
Towards that, it features the \kw{forget} statement
with a \emph{declarative} semantics that magically deletes a quantum variable from the state,
without actual uncomputation happening.
\Cref{fig:forget-statement-iterate} shows how \proc{Iterate}'s uncomputation
can be abstracted this way.
The declarative semantics of \kw{forget} is in general unphysical,
in accordance with the no-deleting theorem \citep{kumar_pati_impossibility_2000}.
However, the authors proved that a physical interpretation exists
in case the forgotten variable was computed \emph{without} any quantum superposition or phase effects.
This condition implies that correct uncomputation could be synthesized,
and is enforced by Silq's effect tracking system.
Unfortunately, due to the expressivity of the language,
the challenge of actually synthesizing the uncomputation has remained open.

\subsubsection{Delaying uncomputation}
One of the keys to efficient uncomputation is to avoid doing it prematurely.
In general, there is more than one point in time when a datum $t$ can be uncomputed.
It might happen, as in \cref{fig:naive-uncomputation},
that the first such point is earlier than a later use of the same datum by another uncomputation.
In this case, uncomputing $t$ at the first opportunity
will only result in recomputing it at that later point,
creating overhead.
If, on the other hand, the datum is kept around as long as it is needed,
then the overhead will be avoided.
For our \proc{Iterate} example
this means accumulating the intermediate iterations $f^{0}(x), \dots, f^{n-1}(x)$ until $y = f^{n}(x)$ is computed,
and then uncomputing them in reverse order.
Delaying uncomputation was understood early on
and it is an essential part of Bennett's constructions \citep{bennett_logical_1973,bennett_timespace_1989} for Turing machines.
For procedural programming languages,
however, there has been only a partial progress so far.

\Citet{james_information_2012} 
interpret what is essentially a classical \kw{forget} instruction as \emph{hiding}:
data that is supposed to be erased is actually discarded into a global garbage dump.
This allowed them to give a reversible semantics of an non-reversible classical language.
The global garbage dump, however, is never uncomputed.
While this is unproblematic for classical programs,
for quantum programs hiding is not equivalent to \kw{forget}.
In fact, \citet{heunen_quantum_2022} noted that hiding alone gives a purification semantics of measurement.
This can be explained with the \emph{principle of implicit measurement} \citep[p. 187]{nielsenchuang}:
The global garbage dump can be assumed to be measured at the end of the execution.
If instead of measurement one needs uncomputation,
then of course the garbage dump has to be uncomputed.
Precisely for that purpose, \citet{protoquipperd} introduced the \emph{garbage monad},
which allows \emph{disposing} data into a local garbage bin that
can be uncomputed when needed.
Delayed uncomputation is done by:
\begin{enumerate*}
\item disposing data instead of uncomputing it right away;
\item uncomputing the whole bin \emph{manually} at the right moment.
\end{enumerate*}

To apply this recipe to \proc{Iterate},
we replace \kw{forget} with \kw{dispose},
obtaining the procedure \proc{Iterate$^\textsf{G}$} in \cref{fig:forget-statement-iterate-garb}.
Note that because the garbage is left uncomputed,
this procedure has a different semantics.
Nonetheless, we can use it to implement \proc{Iterate} efficiently by conjugation:
\[ \proc{Iterate$^G$}(n - 1, f, x, t, \kw{bin}\ g) \; ; \; f(t, y) \; ; \; \proc{Iterate$^{\textsf{G}\dag}$}(n-1, f, x, t, \kw{bin}\dag\ g). \]
Evidently, only twice the work of \proc{Iterate} is done if we count \kw{forget} as costing nothing:
once for the computation and once for the uncomputation.
However, correctness is not guaranteed still.


\subsubsection{Synthesizing uncomputation for circuits}
\Citet{unqomp} gave the first uncomputation synthesis algorithm for arbitrary quantum circuits.
The algorithm transforms a garbage-producing circuit into a garbage-free circuit,
where the garbage qubits have been uncomputed efficiently.
Moreover, it ensures Silq-style correctness by tracking quantum effects of gates.
Also, there is a variant \citep{reqomp} that trades time for space efficiency following \citet{bennett_timespace_1989}.

Even though the method applies to circuits and not to recursive quantum programs,
let us see what it can do for our example.
What we can do is to get a circuit that implements the \proc{Iterate$^\textsf{G}$} procedure for given concrete arguments,
and then run the synthesis algorithm.
The most widespread way for this is to write a \emph{script}, denoted \proc{Iterate$^{\textsf{G}\textsf{C}}$}, that generates the circuit,
e.g., using a library such as Qiskit:
we reinterpret quantum operations as instructions that append gates to the circuit,
procedure calls become calls to scripts,
and \kw{dispose} marks qubits as garbage.
Then, we extend this script with an instruction to run the uncomputation synthesis algorithm of \citeauthor{unqomp}
to obtain the final script \proc{Iterate$^{\textsf{G}\textsf{C}\textsf{V}}$},
which produces an efficient circuit for \proc{Iterate}:
\[ \proc{Iterate$^\textsf{G}$} \mapsto \proc{Iterate$^{\textsf{G}\textsf{C}}$} \mapsto \proc{Iterate$^{\textsf{G}\textsf{C}\textsf{V}}$}. \]

\subsection{Limitations of existing approaches}
We now describe three fundamental orthogonal challenges that need to be addressed.

\subsubsection{Expressivity}
To illustrate the expressivity challenge,
let us consider the \proc{Etareti} procedure in \cref{fig:forget-statement-etareti},
which is defined recursively in terms of its own adjoint.
Such procedures,
where adjoints and uncomputation can be combined freely,
are expressible in modern languages (e.g., Silq).
When we try to execute the sketched uncomputation approach from above,
we encounter a problem at step two,
which translates the procedure to a script \proc{Etareti$^{\textsf{G}\textsf{C}}$}.
We need to interpret the call to the adjoint \proc{Etareti$^{\dag}$} as a call to some script.
The problem here is where does this script actually come from.
Ideally, we would invoke the script \proc{Etareti$^{\dag\textsf{G}\textsf{C}}$},
but to derive it we need an explicit definition of the adjoint.
However, the adjoint is defined only implicitly
and has to be synthesized first.
Unfortunately,
existing methods for adjoint synthesis do not support programs with \kw{forget},
which is a placeholder and not an operation that can be inverted.

An alternative strategy would be to recursively invoke the script \proc{Etareti$^{\textsf{G}\textsf{C}}$} itself,
and then derive a circuit for the supposed call to \proc{Etareti$^{\dag\textsf{G}\textsf{C}}$}.
There are two options,
but neither is satisfactory:
\begin{enumerate}
\item $\proc{Etareti$^{\textsf{G}\textsf{C}}$}(n-1, f, x, t)^\dag$.
  This circuit is not viable because it consumes garbage,
  while we are generating a circuit that produces garbage.
\item $\proc{Etareti$^{\textsf{G}\textsf{C}}$}(n-1, f, x, t)^{\textsf{V}\dag}$.
  This circuit computes the correct result, but it already includes all uncomputation,
  leading to the exponential blowup seen for \proc{Iterate$^\dag$} in \cref{fig:naive-uncomputation-iterate-adj}:
  the circuit for $f^\dag$ will be inlined $2^{n-1}$ times,
  while the one for $f$ will be inlined $2^{n-1} - 1$ times.
\end{enumerate}

This example shows a deeper problem:
there is a vicious circle between adjoint synthesis and uncomputation synthesis.
To synthesize the uncomputation, we need the adjoint.
To synthesize the adjoint, we need the uncomputation.
Therefore, to guarantee efficiency for expressive quantum languages
one has to go beyond existing approaches.

\begin{figure}
  \begin{subfigure}{0.45\textwidth}
    \begin{codebox}
      \Procname{$\proc{Etareti}(n, f, \kw{in}\ x, \kw{out}^\dag\ y)$} \li
      \If $n = 0$:\, $y \gets x \oplus y$ \li
      \Then
      \Else
        \kw{alloc} $t$ \li
        $\proc{Etareti$^{\dag}$}(n-1, f, x, t)$ \li
        $f^\dag(t, y)$ \li
        \kw{forget} $t$ \zi
      \End
    \end{codebox}
    \caption{An alternative implementation of \proc{Iterate$^\dag$}.}
    \label{fig:forget-statement-etareti}
  \end{subfigure}
  \hfill
  \begin{subfigure}{0.45\textwidth}
    \begin{codebox}
      \Procname{$\proc{Etareti$^\textsf{U}$}(n, f, \kw{in}\ x, \kw{out}^\dag\ y)$} \li
      \If $n = 0$:\, $y \gets x \oplus y$ \li
      \Then
      \Else
        \kw{alloc} $t$ \li
        $\proc{Etareti$^{\textsf{U}\dag\textsf{G}}$}(n-1, f, x, t, \kw{bin}\ g)$ \li
        $f^\dag(t, y)$ \li
        $\proc{Etareti$^{\textsf{U}\dag\textsf{G} \dag}$}(n-1, f, x, t, \kw{bin}^\dag\ g)$ \li
        \kw{alloc}$^\dag$ $t$
      \End
    \end{codebox}
    \caption{\proc{Etareti} after uncomputation synthesis.}
  \end{subfigure}
  \hfill
  \begin{subfigure}{0.45\textwidth}
    \begin{codebox}
      \Procname{$\proc{Etareti$^{\textsf{U}\dag}$}(n, f, \kw{in}\ x, \kw{out}\ y)$} \li
      \If $n = 0$:\, $y \gets x \oplus y$ \li
      \Then
      \Else
      \kw{alloc} $t$ \li
      $\proc{Etareti$^{\textsf{U}\dag\textsf{G}}$}(n-1, f, x, t, \kw{bin}\ g)$ \li
      $f(t, y)$ \li
      $\proc{Etareti$^{\textsf{U}\dag\textsf{G}\dag}$}(n-1, f, x, t, \kw{bin}^\dag\ g)$ \li
      \kw{alloc}$^\dag$ $t$
    \End
    \end{codebox}
    \caption{The synthesized adjoint of \proc{Etareti}.}
  \end{subfigure}
  \hfill
  \begin{subfigure}{0.45\textwidth}
    \begin{codebox}
      \Procname{$\proc{Etareti$^{\textsf{U}\dag\textsf{G}}$}(n, f, \kw{in}\ x, \kw{out}\ y, \kw{bin}\ g)$} \li
      \If $n = 0$:\, $y \gets x \oplus y$ \li
      \Then
      \Else
        \kw{alloc} $t$ \li
        $\proc{Etareti}^{\textsf{U}\dag\textsf{G}}(n-1, f, x, t, g)$ \li
        $f^{\textsf{G}}(t, y, g)$ \li
        \kw{dispose} $t$ \kw{in} $g$
      \End
    \end{codebox}
    \caption{Uncomputation erasure of \proc{Etareti$^{\textsf{U}\dag}$}.}
  \end{subfigure}
  \caption{
    Our uncomputation synthesis approach for \proc{Etareti}.
    Here, the \kw{bin} operator binds a garbage bin to a variable,
    and its adjoint \kw{bin$^\dag$} unbinds one,
    uniquely identifying a computation-uncomputation pair.
    The pairing enables \emph{uncomputation erasure}:
    the replacement of uncomputation with \kw{dispose}.
    The resulting procedure accepts a garbage \kw{bin} parameter
    to which temporary variables are disposed;
    the adjoint procedure accepts a \kw{bin$^\dag$} parameter from which garbage can be removed
    using the adjoint operation \kw{dispose$^\dag$}.
  }
  \label{fig:efficient-uncomputation}
\end{figure}

\subsubsection{Modular uncomputation synthesis}
The garbage monad and \citet{unqomp}'s approach (called Unqomp),
discussed above,
facilitate program modularity
by encapsulating garbage.
However, they are not \emph{modular transformations}:
before they can synthesize uncomputation,
they need to inline all subcircuits,
unable to treat them as black-box components.
This leads to two major limitations.
First, synthesis itself is less efficient in case a subcircuit is inlined in many places:
the algorithm will be repeated for every place,
and this can lead to a substantial increase in synthesis running time
in the presence of recursion.
Second, inlining subcircuits can obscure error reporting
because it makes it difficult to track the root cause of an error.
These two considerations make synthesis modularity a desirable property.

\subsubsection{Intermediate representations}
To realize the goal of modular, correct and efficient uncomputation synthesis for expressive quantum programs,
one needs a suitable IR that strikes a balance between three main considerations:
\begin{description}
\item[Expressivity]
  The IR should be expressive enough as a compilation target for the source language
  ensuring that uncomputation details do not leak during lowering.
  For example, when applying Unqomp to \proc{Etareti} in \cref{fig:forget-statement-etareti},
  the call to \proc{Etareti$^\dag$} could be interpreted in the two ways mentioned above,
  both unsatisfactory.
  This would not occur if Unqomp's IR
  supported recursive calls and adjoints.
\item[Simplicity]
  Some IR features complicate matters
  without added benefits.
  For instance, supporting mutable quantum data and even aliasing,
  as in QIR \citep{qir},
  would require tracking old versions of variables,
  as uncomputation of $x$ could depend on an old version of $y$.
  Instead, an SSA based IR makes old versions explicit and is much simpler.
\item[Uncomputation support]
  The IR would ideally include features to support uncomputation.
  For example, it makes sense to distinguish between variable uses that
  ``consume'' the variable from scope,
  such as forgetting a variable,
  and uses that ``conserve'' the variable in scope,
  e.g., in a controlled \proc{not}.
\end{description}

To the best of our knowledge,
no existing IR addresses these concerns simultaneously.

\subsection{Our contributions}

We now discuss our main contributions.

\paragraph{Intermediate representation}
We first introduce a novel intermediate representation (\cref{sec:hqir})
that is designed to address the three concerns discussed above.
Like QIRO \citep{qssa} and QSSA \citep{qssa},
our IR is in SSA form, supports recursion, as well as both quantum and classical computation.
This makes it expressive enough for a wide variety of programs.
Additionally, borrows two features from Silq that help with uncomputation:
\begin{enumerate}
\item It tracks quantum effects to
  determine whether a variable can be correctly uncomputed.
\item It distinguishes between ``consumed'' and ``conserved'' function arguments.
\end{enumerate}

\paragraph{Algorithm}
We then introduce a modular algorithm (\cref{sec:implementation}) for synthesizing correct and efficient uncomputation
even when adjoints and uncomputation are mixed.
The key idea is to extend the transformation $f \mapsto f^{\textsf{G}}$
to replace not only \kw{forget} with \kw{dispose},
but also synthesized uncomputation.
We call this \emph{uncomputation erasure}
as uncomputation is erased,
and the temporary variables are thrown in the bin instead.
The vicious circle between adjoints and uncomputation is broken by synthesizing uncomputation first, and adjoints later.
Then uncomputation erasure can be applied whenever necessary
in order to ensure efficiency.
The approach is illustrated in \cref{fig:efficient-uncomputation}.

\paragraph{Implementation and evaluation}
We have built an end-to-end implementation of our approach consisting of:
a complete implementation of our IR and the uncomputation synthesis algorithm
along with a translation from Silq to the IR
and a code generator from the IR to low-level circuits.
This enables us, for the first time, to handle an expressive fragment of Silq
that includes automatic uncomputation,
reaching a major milestone towards a full compiler.
Further, we have extensively evaluated our system vs. Unqomp on a variety of benchmarks.
The results indicate that we can achieve performance on par with the state of the art,
demonstrating that one can benefit from the greater expressivity and safety offered by high-level languages
without sacrificing efficiency.

%% file: background.tex
\section{Background}
In this section we will introduce the fundamental concepts of quantum computation necessary to understand our work. We refer readers wishing to gain an in-depth understanding of quantum computation to the excellent introduction by \citet{nielsenchuang}.

\subsection{Quantum states}
A quantum computation transforms a quantum state.
\paragraph{One qubit} Quantum information is usually represented in terms of qubits. Like a classical bit, a qubit can be in one of the computational basis states $\ket{0}$ or $\ket{1}$. In contrast to a classical bit, any superposition (complex linear combination) $\varphi=\alpha\ket{0}+\beta\ket{1}$ with $\abs{\alpha}^2+\abs{\beta}^2=1$ is a valid state for a qubit. It is in this sense that a qubit is often thought to be both $0$ and $1$ at the same time. More formally, qubit states are unit vectors in the two-dimensional Hilbert space $\mathcal{H}(\{0,1\})$ spanned by the basis $\{\ket{0},\ket{1}\}$. (In finite dimensions, a Hilbert space is just a complex vector space with an inner product.)

\paragraph{Two qubits} For a pair of qubits, the computational basis states represent all four combinations of two-bit values: $\ket{00}$, $\ket{01}$, $\ket{10}$ and $\ket{11}$. Like in the single-qubit case, any superposition $\varphi=\alpha\ket{00}+\beta\ket{01}+\gamma\ket{10}+\delta\ket{11}$ with $\abs{\alpha}^2+\abs{\beta}^2+\abs{\gamma}^2+\abs{\delta}^2=1$ is a valid state for the two-qubit system.

\paragraph{More qubits} This naturally generalizes to the case of arbitrarily many qubits: An $n$-qubit system has $2^n$ computational basis states representing all combinations of $n$-bit values, and states of the system are given by superpositions of those basis states. More formally, $n$-qubit states are unit vectors in the Hilbert space $\mathcal{H}(\{0,1\}^n)$, where each element of $\{0,1\}^n$ induces one basis vector.

\paragraph{Tensor products} In general, we use the tensor product $\otimes$ to build composite quantum states. The tensor product is a bilinear mapping. For example, the tensor product of the two single-qubit states $\varphi=\alpha\ket{0}+\beta\ket{1}$ and $\varphi'=\alpha'\ket{0}+\beta'\ket{1}$ is given by
\[\varphi\otimes\varphi'=(\alpha\ket{0}+\beta\ket{1})\cdot(\alpha'\ket{0}+\beta'\ket{1})=\alpha\alpha'(\ket{0}\otimes\ket{0})+\alpha\beta'(\ket{0}\otimes\ket{1})+\beta\alpha'(\ket{1}\otimes\ket{0})+\beta\beta'(\ket{1}\otimes\ket{1}).\]
We now simply identify the tensor product $\ket{b}\otimes\ket{b'}$ of computational basis states with the computational basis state $\ket{bb'}$. In this way, we obtain the two-qubit state
\[\varphi\otimes\varphi'=\alpha\alpha'\ket{00}+\alpha\beta'\ket{01}+\beta\alpha'\ket{10}+\beta\beta'\ket{11}.\]
The valid quantum states of the composite system are always superpositions of tensor products of valid states of the two components of the system.

\paragraph{Entanglement} There are composite quantum states that cannot be written as a tensor product of component states. A well-known example is the two-qubit state $\psi=\frac1{\sqrt2}(\ket{01}-\ket{10})$. We can write this state as a superposition of tensor products, namely $\psi=\frac1{\sqrt2}(\ket{1}\otimes\ket{0})-\frac1{\sqrt2}(\ket{0}\otimes\ket{1})$. However, we can show that any tensor product $\varphi\otimes\varphi'$ of single-qubit states cannot be equal to $\psi$. (Consider the expression obtained in the previous paragraph, we would need to simultaneously have $\alpha\alpha'=0$, $\alpha\beta'\neq0$, $\beta\alpha'\neq0$, and $\beta\beta'=0$. This is a contradiction.)

\paragraph{Measurement} To get classical results from a quantum computation, we need to measure the state. If we measure a single-qubit system that is in the state $\varphi=\alpha\ket{0}+\beta\ket{1}$, we will get the value $0$ with probability $\abs{\alpha}^2$ and the value $1$ with probability $\abs{\beta}^2$. Furthermore, the state of the qubit collapses to either $\ket{0}$ or $\ket{1}$ accordingly.

In general, we might measure only one component of a quantum system with multiple components. For example, consider a two-qubit state $\psi$. We want to measure the first component. We first write $\psi$ in the form
\[\psi=\alpha\ket{0}\otimes(\alpha_0\ket{0}+\beta_0\ket{1})+\beta\ket{1}\otimes(\alpha_1\ket{0}+\beta_1\ket{1}),\text{ }\abs{\alpha}^2+\abs{\beta}^2=1, \abs{\alpha_0}^2+\abs{\beta_0}^2=1, \abs{\alpha_1}^2+\abs{\beta_1}^2=1.\]

Such a decomposition always exists and is unique up to relative phases (i.e., a factor of unit magnitude) between $\alpha$ and $\alpha_0, \beta_0$ as well as between $\beta$ and $\alpha_1, \beta_1$. (Group two-qubit basis states by the value they assign to the first bit, pick $\alpha$ and $\beta$ such that they have the same squared magnitude as the sum of squared magnitudes of the coefficients in the respective group.)

After measurement, with probability $\abs{\alpha}^2$ we get result $0$ and the state collapses into $\ket{0}\otimes(\alpha_0\ket{0}+\beta_0\ket{1})$, and with probability $\abs{\beta}^2$ we get result $1$ and the state collapses into $\ket{1}\otimes(\alpha_0\ket{0}+\beta_0\ket{1})$. Note that the resulting states are unique only up to a global phase (i.e., a factor of unit magnitude). This does not bother us because all possible results represent identical physical situations that cannot be distinguished using further measurements.

For example, let $\psi=\frac1{\sqrt2}(\ket{01}-\ket{10})$ be the maximally-entangled state from above. Recall that we can write it as
\[\psi=\frac1{\sqrt2}\ket{0}\otimes\ket{1}-\frac1{\sqrt2}\ket{1}\otimes\ket{0}.\ (\text{I.e., }\alpha=\frac1{\sqrt2}, \beta=-\frac1{\sqrt2}, \alpha_1=\beta_0=1, \alpha_0=\beta_1=0.)\]
Therefore, after measuring the first qubit, we either get result $0$ and state $\ket{01}$ or result $1$ and state $\ket{10}$, both with equal probability.

\subsection{Quantum circuits}
\paragraph{Computation} Quantum computations are often expressed as quantum circuits, representing a transformation on multiple qubits.
Quantum circuits on $n$ qubits are visualized as $n$ horizontal lines, called \emph{wires}. Quantum \emph{gates} operate on one or several wires at a time, transforming the qubits as they pass the wires from left to right.

\newcommand\drawcx{%
  \scalebox{0.5}{
  \begin{quantikz}
      &\ctrl{1}&\push{}\\
      &\targ{}&\push{}\\
  \end{quantikz}
  }%
}

\begin{wrapfigure}[10]{r}{0.3\textwidth}
\begin{quantikz}
  &\gate{H} & \gate[2]{CX} & \gate{Z} & \push{}\\
  &\gate{X} &              & \phantomgate{Z}& \push{}\\
\end{quantikz}
\caption{An example of a quantum circuit. CX is often written \protect\drawcx.}
\end{wrapfigure}
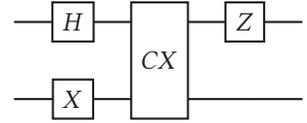

For example, the $2$-qubit quantum circuit on the right uses the single-qubit gates \verb|H|, \verb|X| and \verb|Z|, and the two-qubit gate \verb|CX|. All of those gates apply unitary linear operations to the respective qubits:

\begin{itemize}
\item \verb|H| is the Hadamard transform. It maps $\ket{0}$ to $\frac1{\sqrt2}(\ket{0}+\ket{1})$ and $\ket{1}$ to $\frac1{\sqrt2}(\ket{0}-\ket{1})$. This gate is self-inverse: Applying it twice to the same qubit results in the identity.
\item \verb|X| is the not gate. It maps $\ket{0}$ to $\ket{1}$ and $\ket{1}$ to $\ket{0}$.
\item \verb|CX| is the controlled not gate. It maps $\ket{ab}$ to $\ket{a}\otimes\ket{a\oplus b}$, where $\oplus$ is the XOR operation.
\item \verb|Z| is the controlled phase flip gate. It maps $\ket{0}$ to $\ket{0}$ and $\ket{1}$ to $-\ket{1}$.
\end{itemize}

For our purposes, it is sufficient to write down a sequence of gates and the indices of the qubits they operate on. This fully specifies the operation of a quantum circuit. For example, for the illustrated circuit, we could write \verb|H 0; X 1; CX 0 1; Z 0|, where the top qubit has index $0$ and the bottom qubit has index $1$. This specification has some redundant ordering information. For example, the order of \verb|H 0| and \verb|X 1| does not matter, as they operate on independent qubits.

We now show the operation of our example quantum circuit on the initial state $\ket{00}$, where qubit $0$ is on the left and qubit $1$ is on the right:
\begin{itemize}
\item Initial state: $\ket{00}$.
\item \verb|H 0|: $(\text{H}\ket{0})\otimes\ket{0}=\frac1{\sqrt2}(\ket{0}+\ket{1})\otimes\ket{0}=\frac1{\sqrt2}(\ket{00}+\ket{10})$.
\item \verb|X 1|: $\frac1{\sqrt2}(\ket{0}\otimes(\text{X}\ket{0})+\ket{1}\otimes(\text{X}\ket{0}))=\frac1{\sqrt2}(\ket{01}+\ket{11})$.
\item \verb|CX 0 1|: $\frac1{\sqrt2}(\text{CX}\ket{01}+\text{CX}\ket{11})=\frac1{\sqrt2}(\ket{01}+\ket{10})$.
\item \verb|Z 0|: $\frac1{\sqrt2}((\text{Z}\ket{0})\otimes\ket{1}+(\text{Z}\ket{1})\otimes\ket{0})=\frac1{\sqrt2}(\ket{01}-\ket{10})$.
\end{itemize}
Therefore, our computation produced the entangled state we encountered in earlier examples.

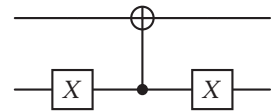
\begin{wrapfigure}[8]{r}{0.3\textwidth}
\begin{quantikz}
  &\phantomgate{s}&\targ{} &\phantomgate{s}&\push{}\\
  &\gate{X}       &\ctrl{-1}&\gate{X}&\push{}\\
\end{quantikz}
\caption{Uncomputation circuit.}
\end{wrapfigure}
\paragraph{Uncomputation} Assume that we are done with qubit $0$ and would like to free it up for further computations by resetting it back into the $\ket{0}$ state. A naive way to achieve this is to measure qubit $0$ and to apply a $\verb|X 0|$ gate if we measure value $1$. However, because qubit $0$ is entangled with qubit $1$, doing so also collapses the state of qubit $1$:
\begin{itemize}
\item Initial state: $\frac1{\sqrt2}(\ket{01}-\ket{10})$.
\item \verb|M 0|: We measure qubit $0$.
  \begin{itemize}
  \item Result $0$, state collapses to $\ket{01}$: Done.
  \item Result $1$, state collapses to $\ket{10}$:
    \begin{itemize}
    \item \verb|X 0|: $(\text{X}\ket{1})\otimes\ket{0}=\ket{00}$. Done.
    \end{itemize}
  \end{itemize}
\end{itemize}

\noindent I.e., we end up in state $\ket{01}$ or $\ket{00}$, each with equal probability. However, we instead would like to end up in the superposed state $\frac1{\sqrt2}(\ket{01}-\ket{00})=\ket{0}\otimes\frac1{\sqrt2}(\ket{1}-\ket{0})$, leaving intact the superposition of the second qubit. The circuit \verb|X 1; CX 1 0; X 1| achieves this:
\begin{itemize}
\item State: $\frac1{\sqrt2}(\ket{01}-\ket{10})$.
\item \verb|X 1|: $\frac1{\sqrt2}(\ket{00}-\ket{11})$.
\item \verb|CX 1 0|: $\frac1{\sqrt2}(\ket{00}-\ket{01})$.
\item \verb|X 1|: $\frac1{\sqrt2}(\ket{01}-\ket{00})=\ket{0}\otimes\frac1{\sqrt2}(\ket{1}-\ket{0})$.
\end{itemize}
I.e., qubit $0$ has been successfully \emph{uncomputed}. In this work, we aim to automatically synthesize uncomputation.

%% file: language.tex
\section{Intermediate representation} \label{sec:hqir}
In this section we introduce our intermediate representation which, as discussed in \cref{sec:intro}, aims to balance expressiveness, simplicity and uncomputation support. We first present the basics of the IR and two features needed for uncomputation synthesis: quantum effect annotations and the distinction between conserved and consumed arguments. We then highlight a variety of features useful to compile expressive quantum programming languages: dynamic qubit allocation, automatic adjoint synthesis, controlled application, classical computation, and multi-qubit values.

The complete syntax of our intermediate representation is shown in \cref{fig:syntax}.
\input{lang-syntax}

\subsection{Functions, variables, and effects}
Each function consists of a list of parameters, a body, and a list of return values.
A body is simply a sequence of statements (there is no block-structure).
Quantum variables denote single qubits (classical and multi-qubit variables are discussed in \cref{sec:hqir-classical,sec:hqir-ext}).
The only kind of statement is a combined variable definition/function call.
Multiple variables can be defined from the results of a single function call.
Arguments between parenthesis are ``consumed'':
the variable is removed from scope.
Ours is an SSA IR, thus each variable can be defined at most once.

To reason about uncomputation correctness, we require that statements are (correctly) annotated with effects, some borrowed from the programming language Silq, but now realized in an IR setting:
\begin{enumerate}
\item $p$ (pure) -- This is the weakest effect. It indicates that the operation consists of pure reversible classical computation applied to qubits, i.e., it maps basis states to basis states (corresponding to the notion of \texttt{qfree} in Silq). Operations that may introduce or destroy superposition or may affect relative or global phase are not permitted. As shown in Silq and Unqomp, this restriction is necessary in order to be able to guarantee uncomputation safety.

\item $q$ (quantum) -- This effect indicates that the operation does not perform measurement (corresponding to the notion of \texttt{mfree} in Silq). However, it is allowed to have any reversible quantum effect, including superposition, relative phase, and global phase, but excluding measurement. We only support adjoint synthesis on functions that have this effect or weaker.

\item $m$ (measure) -- This effect is given to operations that perform measurements. They are irreversible, automatic adjoint synthesis is not supported, and their results cannot be uncomputed.
\end{enumerate}

With this in mind, we can take a look at our first example, a function that creates an EPR pair:

\begin{displaymath}
  \FVar{EPR}(\QVar{a}, \QVar{b}) \coloneq
  \left\{
  \begin{array}{rlrl}
    \QVar{a'}                              & \coloneq_q & \HadamardOp(\QVar{a}) & \\
    \QVar{a''}, \QVar{b'}                  & \coloneq_p & \FVar{CX}(\QVar{a'}, \QVar{b}) &
  \end{array}
  \right\}
  \triangleright \QVar{a''}, \QVar{b'}
\end{displaymath}

Assuming the initial state is $\ket{a b}=\ket{00}$,
after applying the Hadamard gate on $\QVar{a}$, we get $\ket{a' b} = \frac{\ket{00} + \ket{10}}{\sqrt{2}}$.
Then we apply a controlled X gate on $\QVar{b}$ with $\QVar{a'}$ as control,
meaning that the final state is $\ket{\Phi^+} = \frac{\ket{00} + \ket{11}}{\sqrt{2}}$.

\subsection{Conserved use of values}
Many operations take parameters that they return unchanged in the computational basis.
One such example is the $\FVar{CX}$ gate, for which the first parameter is a control
which determines whether an $\FVar{X}$ gate is to be applied to the second parameter.
We call such parameters ``conserved''.
They are not consumed (or returned) by the function and are denoted using brackets.
Our built-in $\FVar{CX}$ gate can now take its control as conserved,
and we update the EPR example accordingly:

\begin{displaymath}
  \FVar{EPR}(\QVar{a}, \QVar{b}) \coloneq
  \left\{
  \begin{array}{rlrl}
    \QVar{a'}                              & \coloneq_q & \HadamardOp(\QVar{a}) & \\
    \QVar{b'}                              & \coloneq_p & \FVar{CX}[\QVar{a'}](\QVar{b}) &
  \end{array}
  \right\}
  \triangleright \QVar{a'}, \QVar{b'}
\end{displaymath}

We use a definition identical to that of Unqomp.
Assuming we have variables $a_1, \ldots, a_n, b_1, \ldots, b_m$
and an operation that conserves $a_1, \ldots, a_n$,
we consider a decomposition of the initial state $\psi \in \hilbert{\set{0,1}^{n+m}}$,
based on the assignment $i \in \set{0,1}^n$ of $a_1, \ldots, a_n$
into separate sub-states $\psi_i \in \hilbert{\set{0,1}^m}$ of $b_1, \ldots, b_m$.
When the resulting quantum state $\psi'$ after the operation is decomposed the same way,
the magnitudes for each $i$ must remain the same:

\begin{gather*}
\psi = \sum_{i \in \set{0,1}^n} \ket{i} \otimes \psi_i
\qquad\to\qquad
\psi' = \sum_{i \in \set{0,1}^n} \ket{i} \otimes \psi'_i \\
\forall i \cdot \abs{\psi_i} = \abs{\psi'_i}
\end{gather*}

The mapping $\psi_i \to \psi'_i$
may be an arbitrary transformation
and may depend on $i$.

We also note that conserved uses of a single variable commute,
so multiple conserved uses do not induce dependencies.
However, it is required that all conserved uses of a variable
happen after the variable is produced, but before it is consumed.
The ability to keep variables unmodified and in scope is critical for uncomputation,
as demonstrated in \cref{sec:hqir-forget} and further elaborated in \cref{sec:uncomp-condition}.

\subsection{Allocation}
We provide a mechanism to encapsulate the allocation of qubits inside functions. While some languages use a stack discipline for allocation (e.g. Q\#), we take a more general approach similar to Quipper and introduce allocation operations. This makes it possible to define functions that perform computation without requiring the caller to preallocate space for the result. The allocation operation $\NewOp{v}$ for $v \in \set{0,1}$ take no arguments and return one qubit.

We can now update our EPR pair example, factoring out a function for creating an entangled copy (note, a copy in the Z basis, not a clone of the quantum state):

\begin{displaymath}
  \DupOp[\QVar{a}] \coloneq
  \left\{
  \begin{array}{rlrl}
    \QVar{b}  & \coloneq_p & \NewOp{0} & \\
    \QVar{b'} & \coloneq_p & \FVar{CX}[\QVar{a}](\QVar{b}) &
  \end{array}
  \right\}
  \triangleright \QVar{b'}
\end{displaymath}

\begin{displaymath}
  \FVar{EPR}() \coloneq
  \left\{
  \begin{array}{rlrl}
    \QVar{a}  & \coloneq_p & \NewOp{0} & \\
    \QVar{a'} & \coloneq_q & \HadamardOp(\QVar{a}) & \\
    \QVar{b}  & \coloneq_p & \DupOp[\QVar{a'}] &
  \end{array}
  \right\}
  \triangleright \QVar{a'}, \QVar{b}
\end{displaymath}

\subsection{Adjoints}
Our representation also supports adjoint function calls. In the adjoint of the function, the roles of the produced variables (left of $\coloneq$) and the consumed variables (inside $()$) are swapped.
The conserved variables (inside $[]$) remain the same, as on the circuit level they are both inputs and outputs. The example below duplicates $a$ into $b$ and then unduplicates $a$ using $b$, implementing an inefficient version of the identity function:

\begin{displaymath}
  \FVar{slow\_id}(\QVar{a}) \coloneq
  \left\{
  \begin{array}{rlrl}
    \QVar{b} & \coloneq_p & \DupOp[\QVar{a}] & \\
             & \coloneq_p & \DupOp^\dagger[\QVar{b}](\QVar{a}) &
  \end{array}
  \right\}
  \triangleright \QVar{b}
\end{displaymath}

The adjoint of allocation $\FVar{new}_{v}^\dagger$ naturally corresponds to deallocation, so we will denote it as $\UnNewOp{v}$ for better readability. Similarly we will use $\UnDupOp$ to denote $\DupOp^\dagger$. Note that when the adjoint of a function is applied incorrectly, e.g., when $\UnNewOp{v}$ is used to deallocate a qubit whose value is not $v$, or when $\UnDupOp[\QVar{b}](\QVar{a})$ is called when $a \oplus b$ may be nonzero, the behavior is undefined.

\subsection{Controlled application}
Many quantum languages and libraries support controlled application of gates and subcircuits.
In our representation, we permit control modifiers on all operations that do not perform measurement (effect $< m$).
For example, the Pauli-$Z$ gate can be expressed as controlled global phase:

\begin{displaymath}
  \FVar{Z}[\QVar{a}] \coloneq
  \left\{
  \begin{array}{rlrl}
    & \coloneq_q & \FVar{phase}_{\pi} & \opIf \QVar{a}
  \end{array}
  \right\}
\end{displaymath}

Controlled application typically means that
the operation is treated as the identity when its condition does not hold.
However, in our IR this cannot be done directly,
as we allow operations with incompatible conserved and produced variables (e.g. $\DupOp$ has outputs but no inputs).
Instead, we introduce a pair of operations:
``distribute'', denoted $\UnSelOp$, that splits one variable into two based on a control,
and ``select'', denoted $\SelOp$, that merges two variables back into one.

When $\UnSelOp$ is applied on a variable, the conserved argument determines
which of the two outputs will be given the value.
The other output will be considered to be undefined.

Similarly, $\SelOp$ is applied on two variables,
where one of them must have a value and the other one must be undefined,
and which one has a value must be correctly identified by the conserved argument.

When the condition of a statement is true,
it is executed, all variables it uses must be defined.
However, if its condition is false, it is not executed,
and all variables it consumes and produces must be undefined.
In general, the condition of a statement can be any conjunctive clause.
In our implementation we ensure the program is valid
using syntactic equivalence (for consumed and produced variables)
and syntactic implication (for conserved variables).

The combination of $\SelOp$ and $\UnSelOp$
lets us keep the SSA form of the IR, and preserves the property that each
variable is consumed once (SSA in reverse).

For example, here is how $\FVar{CX}$ can be expressed as a controlled application of $\FVar{X}$:

\begin{displaymath}
  \FVar{CX}[\QVar{a}](\QVar{b}) \coloneq
  \left\{
  \begin{array}{rlrl}
    \QVar{b_0}, \QVar{b_1} & \coloneq_p & \UnSelOp[\QVar{a}](\QVar{b}) & \\
    \QVar{b_1'}           & \coloneq_p & \FVar{X}(\QVar{b_1}) & \opIf \QVar{a} \\
    \QVar{b'}             & \coloneq_p & \SelOp[\QVar{a}](\QVar{b_0}, \QVar{b_1'}) & \\
  \end{array}
  \right\}
  \triangleright \QVar{b'}
\end{displaymath}

Suppose we start in a state
\[\ket{a b} = \alpha_0\ket{00} + \alpha_1\ket{01} + \alpha_2\ket{10} + \alpha_3\ket{10}\]

We split $b$ based on $a$ into $b_0$ and $b_1$:
\[\ket{a b_0 b_1} = \alpha_0\ket{0 0 \bot} + \alpha_1\ket{0 1 \bot} + \alpha_2\ket{1 \bot 0} + \alpha_3\ket{1 \bot 1}\]

Then when $a=1$ we apply $X$ to $b_1$, producing $b_1'$:
\[\ket{a b_0 b_1'} = \alpha_0\ket{0 0 \bot} + \alpha_1\ket{0 1 \bot} + \alpha_2\ket{1 \bot 1} + \alpha_3\ket{1 \bot 0}\]

Finally, we merge $b_0$ and $b_1'$:
\[\ket{a b'} = \alpha_0\ket{0 0} + \alpha_1\ket{0 1} + \alpha_2\ket{1 1} + \alpha_3\ket{1 0}\]

Note that while we define controlled application via the undefined value $\bot$,
the program correctness requirements stated above mean
that implementations do not need an explicit representation of $\bot$,
and may even allocate multiple variables to the same qubits
if it is known that at most one of them is defined.

\subsection{Automatic uncomputation}\label{sec:hqir-forget}
We define $\ForgetOp$ to be a special form of deallocation that does not need to be told the value of the variable it is deallocating. On Hilbert spaces this corresponds to erasing the variable from the labels of the basis vectors.

For example, we can implement the majority function $\mathit{maj}(a, b, c) = [a + b + c \geq 2]$
by first computing $x = a \oplus b$,
then the result $r = \begin{cases}b & \text{if $x=0$} \\ c & \text{if $x=1$}\end{cases}$,
and finally using $\ForgetOp$ to clean up $x$.

\begin{displaymath}
  \FVar{maj}[\QVar{a}, \QVar{b}, \QVar{c}] \coloneq
  \left\{
  \begin{array}{rlrl}
    \QVar{t}  & \coloneq_p & \DupOp[\QVar{a}] & \\
    \QVar{x}  & \coloneq_p & \FVar{CX}[\QVar{b}](\QVar{t}) & \\
    \QVar{r_0} & \coloneq_p & \DupOp[\QVar{b}] & \opIf{\overline{\QVar{x}}} \\
    \QVar{r_1} & \coloneq_p & \DupOp[\QVar{c}] & \opIf{\QVar{x}} \\
    \QVar{r}  & \coloneq_p & \SelOp[\QVar{x}](\QVar{r_0}, \QVar{r_1}) & \\
              & \coloneq_p & \ForgetOp(\QVar{x}) &
  \end{array}
  \right\}
  \triangleright \QVar{r}
\end{displaymath}

Suppose we apply $\FVar{maj}$ to the state

\begin{align*}
\ket{abc} &= \frac{\ket{0}+\ket{1}}{\sqrt{2}} \otimes \frac{\ket{0}-\ket{1}}{\sqrt{2}} \otimes \ket{1} \\
          &= \frac{1}{2} \ket{001} - \frac{1}{2} \ket{011} + \frac{1}{2} \ket{101} - \frac{1}{2} \ket{111}
\end{align*}

After computing $x$ and $r$, we have
\begin{align*}
\ket{abcxr} &= \frac{1}{2} \ket{00100} - \frac{1}{2} \ket{01111} + \frac{1}{2} \ket{10111} - \frac{1}{2} \ket{11101}
\end{align*}

Then $\ForgetOp(x)$ corresponds to erasing $x$:

\begin{align*}
\ket{abcr} &= \frac{1}{2} \ket{0010} - \frac{1}{2} \ket{0111} + \frac{1}{2} \ket{1011} - \frac{1}{2} \ket{1111}
\end{align*}

In this case we know that $\ForgetOp(x)$ is realizable, because $x$ is computed from $\QVar{b}$ and $\QVar{a}$ (the latter is used to compute the temporary $\QVar{t})$, and they are both in scope.
However, in some cases $\ForgetOp$ can unphysical, as it can collapse different non-zero-amplitude basis vectors. In the example below, after the Hadamard gate $\HadamardOp$ we are in a state $\frac{\ket{0} + \ket{1}}{\sqrt{2}}$, and after the $\ForgetOp(\QVar{a'})$ we would be in $\frac{\ket{\cdot} + \ket{\cdot}}{\sqrt{2}} = \sqrt{2}\ket{\cdot}$, meaning that the transformation is not unitary.

\begin{displaymath}
  \FVar{BAD}() \coloneq
  \left\{
  \begin{array}{rlrl}
    \QVar{a}  & \coloneq_p & \NewOp{0} & \\
    \QVar{a'} & \coloneq_q & \HadamardOp(\QVar{a}) & \\
              & \coloneq_p & \ForgetOp(\QVar{a'}) &
  \end{array}
  \right\}
\end{displaymath}

Furthermore, in some cases it may be safe to $\ForgetOp$ a variable, but doing so would be computationally difficult. In the following example, assuming $\FVar{f}_\textit{inj}$ implements an injective function, the $\ForgetOp$ would be possible, as $\QVar{x}$ is a function of $\QVar{y}$. However, implementing it would be equivalent to finding an inverse of the $\FVar{f}_\textit{inj}$, which may not be easy.

\begin{displaymath}
  \FVar{HARD}(\QVar{x}) \coloneq
  \left\{
  \begin{array}{rlrl}
    \QVar{y} & \coloneq_p & \FVar{f}_\textit{inj}[\QVar{x}] & \\
             & \coloneq_p & \ForgetOp(\QVar{x})
  \end{array}
  \right\}
  \triangleright \QVar{y}
\end{displaymath}

In \cref{sec:uncomp} we discuss our algorithm for replacing $\ForgetOp$ with explicit uncomputation and the cases it supports.

\subsection{Classical variables} \label{sec:hqir-classical}
We also extend our representation with support for classical variables, which we denote with subscript \texttt{c}. Classical variables cannot be consumed and may be implicitly discarded at the end of the function. They are returned by the $\MeasureOp$ built-in and can be used as controls. For example, here is how one could implement quantum teleportation \cite{quantum_teleportation} given $(\QVar{a}, \QVar{b}) \coloneq_q \FVar{EPR}()$:

\begin{displaymath}
  \FVar{send}(\QVar{a}, \QVar{x}) \coloneq
  \left\{
  \begin{array}{rlrl}
    \QVar{a_0}, \QVar{a_1} & \coloneq_p & \UnSelOp[\QVar{x}](\QVar{a}) & \\
    \QVar{a_1'}           & \coloneq_p & \FVar{X}(\QVar{a_1}) & \opIf \QVar{x} \\
    \QVar{a'}            & \coloneq_p & \SelOp[\QVar{x}](\QVar{a_0}, \QVar{a_1'}) & \\
    \CVar{a}             & \coloneq_m & \MeasureOp(\QVar{a'}) & \\
    \QVar{x'}            & \coloneq_q & \HadamardOp(\QVar{x}) & \\
    \CVar{x}             & \coloneq_m & \MeasureOp(\QVar{x'}) &
  \end{array}
  \right\}
  \triangleright \CVar{a}, \CVar{x}
\end{displaymath}

\begin{displaymath}
  \FVar{recv}[\CVar{a}, \CVar{x}](\QVar{b}) \coloneq
  \left\{
  \begin{array}{rlrl}
    \QVar{b_0}, \QVar{b_1} & \coloneq_p & \UnSelOp[\CVar{a}](\QVar{b}) & \\
    \QVar{b_1'}           & \coloneq_p & \FVar{X}(\QVar{b_1}) & \opIf \CVar{a} \\
    \QVar{y}             & \coloneq_p & \SelOp[\CVar{a}](\QVar{b_0}, \QVar{b_1'}) & \\
                         & \coloneq_q & \PhaseOp{\pi} & \opIf \CVar{x} \land \QVar{y}
  \end{array}
  \right\}
  \triangleright \QVar{y}
\end{displaymath}

To make reasoning about programs easier, we require that all classical variables be marked as such. In general, it is allowed for classical variables to have composite types, assuming the necessary built-in functions are provided.

With the support for classical computation and conditionals, it is also possible for functions to be recursive.

\subsection{Multi-qubit values} \label{sec:hqir-ext}
Our representation also supports composite values. For example, multi-qubit variables could be used via concatenation ($\CatOp{}$) and its adjoint ($\UnCatOp{}$, partitioning).
Note that in order for adjoint synthesis (\cref{sec:adjoint}) to work, we need to make sure that both $\CatOp{}$ and $\UnCatOp{}$ are given the sizes of the variables they are working on.
For example, we can use these operations to define a function $\FVar{extract}[\CVar{n}, \CVar{i}](a)$ for extracting the $\CVar{i}$-th element of an $\CVar{n}$-element qubit array $a$, returning both the extracted element and the remaining $\CVar{n}-1$ elements:

\begin{displaymath}
  \FVar{extract}[\CVar{n}, \CVar{i}](\QVar{a}) \coloneq
  \left\{
  \begin{array}{rlrl}
    \CVar{j}                     & \coloneq_p & \CVar{n} - \CVar{1} - \CVar{i} & \\
    \QVar{x}, \QVar{y}, \QVar{z} & \coloneq_p & \UnCatOp{3}[\CVar{i}, \CVar{1}, \CVar{j}](\QVar{a}) & \\
    \QVar{a'}                    & \coloneq_p & \CatOp{2}[\CVar{i}, \CVar{j}](\QVar{x}, \QVar{z})
  \end{array}
  \right\}
  \triangleright \QVar{a'}, \QVar{y}
\end{displaymath}

All these features allow the IR to be a compilation target for higher-level languages, including Silq (discussed in \cref{sec:evaluation}) and of course, simpler representations such as circuits.

%% file: lang-syntax.tex
\begin{figure}\small
\begin{align*}
  f \Coloneqq   &\quad \ldots && \text{(functions)} \\
  \QVar{x} \Coloneqq   &\quad \ldots && \text{(quantum variables)} \\
  \CVar{x} \Coloneqq &\quad \ldots && \text{(classical variables)} \\
  o \Coloneqq   &\quad f \;|\; \SelOp \;|\; \UnSelOp \;|\; \ForgetOp && \text{(operations)} \\
  c \Coloneqq   &\quad \CVar{x} \;|\; \overline{\CVar{x}} \;|\; \QVar{x} \;|\; \overline{\QVar{x}} && \text{(conditions)} \\
  e \Coloneqq   &\quad p \;|\; q \;|\; m && \text{(effects)}\\
  s \Coloneqq   &\quad \vec{\CVar{x}}, \vec{\QVar{x}} \coloneq_e o[\vec{\CVar{x}}, \vec{\QVar{x}}](\vec{\QVar{x}}) \opIf \vec{c} && \text{(statements)} \\
  F \Coloneqq   &\quad f[\vec{\CVar{x}}, \vec{\QVar{x}}](\vec{\QVar{x}}) \coloneq \{\vec{s}\} \triangleright \vec{\CVar{x}}, \vec{\QVar{x}} && \text{(function definitions)} \\
  P \Coloneqq   &\quad \vec{F} && \text{(program)}
\end{align*}
\caption{\label{fig:syntax}Syntax}
\end{figure}

%% file: implementation.tex
\section{Algorithms}\label{sec:implementation}
In this section, we present algorithms for uncomputation synthesis as
well as adjoint synthesis. We first show a modular
precondition on the validity of a $\ForgetOp$ statement that can be
checked at the function level. This precondition ensures that
uncomputation synthesis will succeed.

Furthermore, we will argue that uncomputation synthesis correctly
implements $\ForgetOp$ statements in terms of explicit uncomputation,
and that an arbitrary combination of uncomputation and adjoint
synthesis always preserves the asymptotic running time of a program.

\input{impl-uncomp}
\input{impl-adjoint}
\input{impl-garbage}

%% file: impl-uncomp.tex
\subsection{Uncomputation precondition}\label{sec:uncomp-condition}
As mentioned in \cref{sec:hqir-forget}, $\ForgetOp$ statements may not always be realizable.
We define a simple modular precondition that we will show is sufficient for our uncomputation synthesis procedure:

\begin{property}[Forgettable at location] \label{property:forgettable}
  A quantum variable $x$ is \emph{forgettable at a program point $P$} if
  \begin{enumerate}
  \item The variable is defined as the result of a statement $S$ with $\attrib{S}{effect} < q$ mode before point $P$.
  \item Any quantum argument to this statement $S$ is either in scope or forgettable at point $P$.
  \end{enumerate}
\end{property}
\noindent If $x$ is forgettable at a point $P$, the quantum arguments to the call $C$ are called the \emph{dependencies} of $x$.

A variable may be forgettable at some program point even if it has already gone out of scope by that point. Intuitively, a quantum variable $\QVar{x}$ is forgettable at some program point $P$ iff it can be recomputed at point $P$ from available information other than $\QVar{x}$ itself, separately in each basis state.

Consider the function $\FVar{maj}$ from \cref{sec:hqir-forget}. The variable $\QVar{x}$ is forgettable at the point $P$ of the $\ForgetOp(x)$ because (1) it is the result of the pure call $\QVar{x} \coloneq_p \FVar{CX}[\QVar{b}](\QVar{t})$ (i.e., without quantum effects such as phase or superposition) and (2) the quantum argument $\QVar{b}$ to this call is in scope at point $P$ and the quantum argument $\QVar{t}$ is forgettable at point $P$ (though it is not in scope). Quantum argument $\QVar{t}$ is forgettable at point $P$ because (1) it is the result of the pure call $\QVar{t} \coloneq_p \DupOp[\QVar{a}]$ and (2) the quantum argument $\QVar{a}$ to this call is still in scope at point $P$.

\begin{property}[Well-forgotten] \label{property:well-forgotten}
  A function $\FVar{f}$ is \emph{well-forgotten} if for every $\ForgetOp$ statement $S$ in the function $\FVar{f}$, every quantum argument to $S$ is forgettable at the program point of $S$.
\end{property}

It immediately follows that in a pure function with no consumed
arguments, every quantum variable that is not a conserved argument is
always forgettable after its definition, and therefore the function is
well-forgotten. We have (1) because all statements within a pure function
are pure, and (2) by induction because every argument is conserved and
therefore always in scope.

The pure function $\FVar{maj}[\QVar{a}, \QVar{b}, \QVar{c}]$ is hence well-forgotten.
However, the function $\FVar{BAD}()$ from \cref{sec:hqir-forget}
is not well-forgotten, as condition (1) does not hold: the quantum variable
$\QVar{a'}$ is not the result of a pure call (its definition
$\QVar{a'} \coloneq_q \HadamardOp(\QVar{a})$ has a quantum effect
annotation $\coloneq_q$).

\subsection{Uncomputation synthesis}\label{sec:uncomp}
The definition of \cref{property:well-forgotten} and the underlying
intuition suggests a naive implementation of $\ForgetOp(x)$ that
recomputes all dependencies of $x$ that are not already in scope, uses
them to uncompute $x$ using the adjoint of its defining call, and then
uncomputes the dependencies.  Unfortunately, this is likely to result
in a very inefficient program, possibly even with exponential
overhead, as every $\ForgetOp{}$ statement may increase the size of a
program prefix by a constant factor.

\begin{figure}\small
  \input{algo-uncomp}
  \caption{Uncomputation synthesis}
  \label{fig:uncomp}
\end{figure}

Instead, in \cref{fig:uncomp}, we describe an efficient uncomputation synthesis algorithm that
does not perform any redundant recomputation.
We present our algorithm as a single backwards pass over a function.
When a statement $S \simeq \Stmt{\ForgetOp(\QVar{x})}$ is encountered,
we ensure that $\QVar{x}$ is uncomputed using $\proc{Ensure-Uncomputed}(S,x)$
in line 6 of \proc{Synthesize-Uncomputation}.
If $\QVar{x}$ has been previously uncomputed at a later point in the function
(recall that we transform the function backwards), nothing remains to be done
(line 5 of \proc{Ensure-Uncomputed}).
Otherwise, $\QVar{x}$ still has to be uncomputed.

To uncompute $\QVar{x}$, we synthesize the adjoint of the statement that computed $x$ using the call $\proc{Undo-Statement}(S,C)$ in line 7 of \proc{Ensure-Uncomputed}.
Suppose we are undoing $C$, a call to $\FVar{f}$.
If some of the inputs to $C$ are not in scope,
we make them available by forcing their uncomputation to happen later in the function
(lines 4-5 of \proc{Undo-Statement}).
Then, we insert a new call $U$ to $\FVar{f}^\dagger$
(lines 8-9)
and clean up its outputs, if any
(lines 12-13).

The call to $\FVar{f}^\dagger$ consumes exactly the results of the call to $\FVar{f}$.
However, they are already consumed by the function elsewhere,
which we need to resolve.
In the case where they are still in scope,
i.e., the original consumption is after the current point,
we create duplicates that are used for the remainder of the function
(lines 15-20 of \proc{Undo-Statement}).
If they are not in scope,
meaning that their original consumption is before the current point,
we mark them for lifetime extension
(line 22 of \proc{Undo-Statement}),
which will result in a $\DupOp$ being inserted later,
when the original consumer is visited
(lines 10-13 of \proc{Synthesize-Uncomputation}).
In the case where the original consumer is a $\ForgetOp$,
no duplication takes place as the $\ForgetOp$ is removed
(line 7 of \proc{Synthesize-Uncomputation}).

In order to keep the presentation simple, the pseudocode in $\cref{fig:uncomp}$
does not handle conditionals. The complete implementation does --
all we need to do is to define the conditions on the newly synthesized statements.
In $\proc{Undo-Statement}$, the newly-added statement $U$
needs to have the same condition as $C$.
The conditions of the $\DupOp$ and $\UnDupOp$ operations
are the same as the conditions of the statements before which they are inserted,
optionally also conditioned on the control argument if that statement is a $\SelOp$ or $\UnSelOp$.

For now, we completely ignore $\proc{Connect-Garbage}$, which we will elaborate upon in \cref{sec:garbage}.
Consider again our example $\FVar{maj}$ from $\cref{sec:hqir-forget}$.
Uncomputation synthesis on this function results in the following new function:
\begin{displaymath}
  \FVar{maj}^\textsc{U}[\QVar{a}, \QVar{b}, \QVar{c}] \coloneq
  \left\{
  \begin{array}{rlrl}
    \QVar{t}  & \coloneq_p & \DupOp[\QVar{a}] & \\
    \QVar{t''}& \coloneq_p & \DupOp[\QVar{t}] & \\
    \QVar{x}  & \coloneq_p & \FVar{CX}[\QVar{b}](\QVar{t''}) & \\
    \QVar{r_0} & \coloneq_p & \DupOp[\QVar{b}] & \opIf{\overline{\QVar{x}}} \\
    \QVar{r_1} & \coloneq_p & \DupOp[\QVar{c}] & \opIf{\QVar{x}} \\
    \QVar{r}  & \coloneq_p & \SelOp[\QVar{x}](\QVar{r_0}, \QVar{r_1}) & \\
    \QVar{t'} & \coloneq_p & \FVar{CX}^\dagger[\QVar{b}](\QVar{x}) & \\
              & \coloneq_p & \UnDupOp[\QVar{t}](\QVar{t'}) & \\
              & \coloneq_p & \UnDupOp[\QVar{a}](\QVar{t})
  \end{array}
  \right\}
  \triangleright \QVar{r}
\end{displaymath}

The statement $\ForgetOp(x)$ has been replaced with the explicit uncomputation code $\QVar{t'} \coloneq_p \FVar{CX}^\dagger[\QVar{b}](\QVar{x})$, and calls to $\DupOp$ and $\UnDupOp$ have been inserted to ensure that values are in scope when they are needed for uncomputation and then properly disposed of again after they are no longer required.

\paragraph{Explanation} The algorithm initially encounters the $S=\Stmt{\ForgetOp(x)}$ statement and calls the subprocedure $\proc{Ensure-Uncomputed}(S, x)$. We have $C=\id{producer}(x)=\Stmt{\QVar{x} \coloneq_p \FVar{CX}[\QVar{b}](\QVar{t})}$. As $C$ has not been undone yet, we call $\proc{Undo-Statement}(S,C)$.

In order to ensure that we have access to the dependency $\QVar{t}$, the algorithm then calls the subprocedure $\proc{Ensure-Uncomputed}(S,\QVar{t})$, which in turn sets $C'=\id{producer}(t)=\Stmt{\DupOp[\QVar{a}]}$ and calls $\proc{Undo-Statement}(S,C')$. This inserts the statement $\UnDupOp[\QVar{a}](\QVar{t})$ after $S$ (line 9) and further adds $\QVar{t}$ to the set $\id{extended}$ to later extend its lifetime (line 22). Execution then returns to $\proc{Undo-Statement(S,C)}$ at line 6. This inserts the explicit uncomputation code $U=\Stmt{\QVar{t'} \coloneq_p \FVar{CX}^\dagger[\QVar{b}](\QVar{x})}$ after $S$ (line 8) and finally inserts $\UnDupOp[\QVar{t}](\QVar{t'})$ after $U$ to clean up the result $\QVar{t'}$.

Execution then returns to $\proc{Synthesize-Uncomputation}(\FVar{f})$, which first removes $S=\Stmt{\ForgetOp(x)}$ (line 7) and proceeds to only update the $\id{alive}$ variables (line 14) until it encounters the statement $S=\Stmt{\QVar{x} \coloneq_p \FVar{CX}[\QVar{b}](\QVar{t})}$, which consumes the variable $\QVar{t}$ that we had put into the set $\id{extended}$. Lines 12-13 extend the lifetime of $t$ by creating the duplicate $\QVar{t''} \coloneq_p \DupOp[\QVar{t}]$ and making the statement $S$ consume $\QVar{t''}$ instead of $\QVar{t}$. Execution then terminates with the updated function $\FVar{maj}^\textsc{U}$ shown above.

\paragraph{Simplifications} We can further perform basic simplifications to remove redundant $\DupOp$ calls.
For example, $\QVar{t}$ is computed as $\DupOp[\QVar{a}]$ and then uncomputed as $\UnDupOp[\QVar{a}]$,
so it can be replaced with $\QVar{a}$:%
\begin{displaymath}
  \FVar{maj}^\textsc{U}[\QVar{a}, \QVar{b}, \QVar{c}] \coloneq
  \left\{
  \begin{array}{rlrl}
    \QVar{t''}& \coloneq_p & \DupOp[\QVar{a}] & \\
    \QVar{x}  & \coloneq_p & \FVar{CX}[\QVar{b}](\QVar{t''}) & \\
    \QVar{r_0} & \coloneq_p & \DupOp[\QVar{b}] & \opIf{\overline{\QVar{x}}} \\
    \QVar{r_1} & \coloneq_p & \DupOp[\QVar{c}] & \opIf{\QVar{x}} \\
    \QVar{r}  & \coloneq_p & \SelOp[\QVar{x}](\QVar{r_0}, \QVar{r_1}) & \\
    \QVar{t'} & \coloneq_p & \FVar{CX}^\dagger[\QVar{b}](\QVar{x}) & \\
              & \coloneq_p & \UnDupOp[\QVar{a}](\QVar{t'}) & \\
  \end{array}
  \right\}
  \triangleright \QVar{r}
\end{displaymath}

\input{proof-completeness}

A proof that this procedure indeed synthesizes correct uncomputation for all well-forgotten functions
is provided in \cref{appendix:proofs-uncomp}.

%% file: algo-uncomp.tex
\begin{subfigure}[t]{0.55\textwidth}
\begin{codebox}
  \Procname{$\proc{Synthesize-Uncomputation}(\FVar{f})$} \li
    $\id{alive} \gets \attrib{\FVar{f}}{conserved} \cup \attrib{\FVar{f}}{produced}$\li
    $\id{undone} \gets \set{}$\li
    $\id{extended} \gets \set{}$\li
    \For $S \in \id{reversed}(\attrib{\FVar{f}}{statements})$:\li
    \Do
      \If $\Stmt{\ForgetOp(x)} \gets S$:\li
      \Then
        $\proc{Ensure-Uncomputed}(S, x)$\li
        $\proc{Remove}(S)$\li
        \textbf{continue}
      \End\li
      \Comment{fix variables with extended lifetimes}\li
      \For $x \in \attrib{S}{consumed} \cap \id{extended}$:\li
      \Do
        $x' \gets \proc{Fresh-Variable}()$\li
        $\proc{Insert-Before}\left(S, \Stmt{x' \coloneq \DupOp[x]}\right)$\li
        $\proc{Substitute}(S, x \mapsto x')$
      \End\li
      $\id{alive} \gets (\id{alive} \setminus \attrib{S}{produced}) \cup \attrib{S}{consumed}$
    \End
  \End
\end{codebox}%
\begin{codebox}
  \Procname{\textbf{subprocedure }$\proc{Ensure-Uncomputed}(S, x)$}\li
    \Comment{ensure $x$ is in scope at $S$ and uncomputed later}\li
    \textbf{assert} $x \not\in \attrib{\FVar{f}}{conserved} \cup \attrib{\FVar{f}}{consumed}$\li
    $C \gets \id{producer}(x)$\li
    \textbf{assert} $\attrib{C}{effect} < q$\li
    \If $C \not\in \id{undone}$:\li
    \Then
      $\id{undone} \gets \id{undone} \cup \set{C}$\li
      $\proc{Undo-Statement}(S, C)$
    \End
  \End
\end{codebox}
\end{subfigure}%
\begin{subfigure}[t]{0.45\textwidth}
\begin{codebox}
  \Procname{\textbf{subprocedure }$\proc{Undo-Statement}(S, C)$}\li
    $\Stmt{\vec{y} \coloneq \FVar{g}[\vec{c}](\vec{x})} \gets C$\li
    \Comment{ensure all inputs to $C$ are}\li
    \Comment{alive or uncomputed after $S$}\li
    \For $x \in (\set{\vec{c}} \setminus \id{alive}) \cup \set{\vec{x}}$:\li
    \Do
      $\proc{Ensure-Uncomputed}(S, x)$
    \End\li
    \Comment{insert inverse of $C$ right after $S$, before}\li
    \Comment{any newly-synthesized uncomputation}\li
    $U \gets \Stmt{\vec{x}' \coloneq \FVar{g}^\dagger[\vec{c}](\vec{y})}$\quad($\vec{x}'$ fresh variables)\li
    $\proc{Insert-After}(S, U)$\li
    $\proc{Connect-Garbage}(C, U)$\li
    \Comment{clean up the results produced by $U$}\li
    \For $(x, x') \in \id{zip}(\vec{x}, \vec{x}')$:\li
    \Do
      $\proc{Insert-After}\left(U, \Stmt{\UnDupOp[x](x')}\right)$
    \End\li
    \Comment{$\DupOp$ variables consumed after $S$}\li
    \For $y \in \set{\vec{y}} \cap \id{alive}$:\li
    \Do
      \Comment{$y$ is consumed after $S$, so $\DupOp$ it}\li
      \Comment{and use the duplicate after $U$}\li
      $y' \gets \proc{Fresh-Variable}()$\li
      $\proc{Insert-Before}\left(U, \Stmt{y' \coloneq \DupOp[y]}\right)$\li
      $\proc{Substitute}(\id{all-after}(S), y \mapsto y')$
      \End
    \End\li
    \Comment{extend results consumed before $S$}\li
    $\id{extended} \gets \id{extended} \cup (\set{\vec{y}} \setminus \id{alive})$
  \End
\end{codebox}
\end{subfigure}%

%% file: proof-completeness.tex
\paragraph{Completeness}
We show that for any well-forgotten function $\FVar{f}$,
$\proc{Synthesize-Uncomputation}(\FVar{f})$ terminates successfully,
i.e., the assertions in $\proc{Ensure-Uncomputed}$ do not fail.

\paragraph{Preconditions} We will show that for all invocations of $\proc{Ensure-Uncomputed}(S, x)$,
$x$ is not in scope and forgettable at the point after $S$,
and that $\proc{Undo-Statement}(S, x)$ is only called on statements
that produce at least one variable that is forgettable at the point after $S$.

\begin{itemize}
\item When $\proc{Synthesize-Uncomputation}$ calls $\proc{Ensure-Uncomputed}(S, \id{x})$ for some statement $S = \Stmt{\ForgetOp(x)}$ on line 6,
this holds because the function $\FVar{f}$ is well-forgotten, therefore $x$ is forgettable at $S$, and because $x$ is consumed by $S$, therefore not in scope.

\item When $\proc{Ensure-Uncomputed}(S, \id{x})$ calls $\proc{Undo-Statement}(S, producer(\id{x}))$ on line 7,
this holds because $x$ is forgettable at the statement $S$.

\item When $\proc{Undo-Statement}(S, C)$ calls $\proc{Ensure-Uncomputed}(S, \id{x})$ on line 5 for some variable $\id{x}\in (\attrib{C}{conserved} \setminus \id{alive}) \cup \attrib{C}{consumed}$,
  the precondition holds because $C$ produces a variable that is forgettable at $S$, and $\QVar{x}$ is an argument to $C$ that is not in scope, so $\QVar{x}$ is also forgettable, due to the well-forgotten property.
  
Note that $\attrib{C}{consumed} \cap \id{alive} = \set{}$, as consumed variables are removed from scope.

\item Finally, even though $\proc{Undo-Statement}$ modifies the function, it only makes changes after $S$, which does not affect the forgettability of variables at $S$.
\end{itemize}

As in $\proc{Ensure-Uncomputed}(S, \QVar{x})$, the variable $\QVar{x}$ is forgettable and not in scope, it must be a parameter and must be produced by a function without quantum effects. Therefore the assertions hold.

%% file: impl-adjoint.tex
\subsection{Adjoint synthesis}\label{sec:adjoint}
The next transformation we present is adjoint synthesis.
The goal of this transformation is, given a function $\FVar{f}$, to generate its adjoint $\FVar{f}^\dagger$.
Note that for adjoint synthesis to be possible, $\FVar{f}$ must not perform any measurements,
as measurement is usually irreversible and therefore its adjoint cannot be defined in general.

\paragraph{Algorithm}

First, we define the transformation $\proc{Synthesize-Classical}(\FVar{f})$ that,
given some function $\FVar{f}$ with effect $< m$, generates a function $\FVar{f}^\textsc{O}$
that only performs the classical computation of $\FVar{f}$
without any quantum effects,
and only has classical inputs and outputs.
It works by transforming each statement in order.
Classical built-ins are transformed to themselves, and quantum built-ins become no-ops.
We need this step because the potentially irreversible classical computation still has to proceed in forward direction, while we need to reverse the order of the quantum computations.

Next, we define $\proc{Synthesize-Adjoint}(\FVar{f})$ that synthesizes the adjoint of a function $\FVar{f}$.
The result is a function $\FVar{f}^\dagger$,
where the roles of the consumed quantum arguments and produced quantum results have been swapped.
The algorithm is quite simple --
we first synthesize the classical computation, which gives us all classical variables defined in $\FVar{f}$,
and then copy over the adjoints of all quantum statements in reverse order.
\begin{figure}\small
  \input{algo-adjoint}
  \caption{Adjoint synthesis}
  \label{fig:adjoint}
\end{figure}
\cref{fig:adjoint} displays those algorithms in pseudocode.

One interesting example is $\FVar{extract}$ from \cref{sec:hqir-ext},
where we need to perform the classical computation first,
and then reverse only the quantum computation:%
\begin{displaymath}
  \FVar{extract}^\dagger[\CVar{n}, \CVar{i}](\QVar{a'}, \QVar{y}) \coloneq
  \left\{
  \begin{array}{rlrl}
    \multicolumn{3}{l}{\text{\Comment{classical part}}} \\
    \CVar{j}           & \coloneq_p & \CVar{n} - \CVar{1} - \CVar{i} & \\
    \multicolumn{3}{l}{\text{\Comment{quantum part}}} \\
    \QVar{x}, \QVar{z} & \coloneq_p & \UnCatOp{2}[\CVar{i}, \CVar{j}](\QVar{a'}) \\
    \QVar{a}           & \coloneq_p & \CatOp{3}[\CVar{i}, \CVar{1}, \CVar{j}](\QVar{x}, \QVar{y}, \QVar{z})
  \end{array}
  \right\}
  \triangleright \QVar{a}
\end{displaymath}
The function $\FVar{extract}^\dagger[\CVar{n}, \CVar{i}](\QVar{a'}, \QVar{y})$ inserts a qubit $\QVar{y}$ at index $\CVar{i}$ of a the multi-qubit variable $\QVar{a'}$.

%% file: algo-adjoint.tex
\begin{subfigure}[t]{0.45\textwidth}
\begin{codebox}
  \Procname{$\proc{Synthesize-Classical}(\FVar{f})$} \li
    \textbf{assert} $\attrib{\FVar{f}}{effect} < m$\li
    $\FVar{f}^\textsc{O} \gets \proc{New-Function}()$\li
    $\attrib{\FVar{f}^\textsc{O}}{classical-in} \gets \attrib{\FVar{f}}{classical-in}$\li
    $\attrib{\FVar{f}^\textsc{O}}{classical-out} \gets \attrib{\FVar{f}}{classical-out}$\li
    \For $S \in \attrib{\FVar{f}}{statements}$:\li
    \Do
      \If $\attrib{S}{classical-out} \neq \set{}$:\li
      \Then
        $S^\textsf{O} \gets \proc{Make-Classical}(S)$\li
        $\proc{Append}(\FVar{f}^\textsc{O}, S^\textsc{O})$
      \End
    \End\li
    \Return $\FVar{f}^\textsf{O}$
  \End
\end{codebox}

\begin{codebox}
  \Procname{$\proc{Make-Classical}(S)$} \li
    $\Stmt{\vec{\CVar{y}}, \vec{y} \coloneq \FVar{g}[\vec{\CVar{c}}, \vec{c}](\vec{x})} \gets S$\li
    \Return $\Stmt{\vec{\CVar{y}} \coloneq \FVar{g}^\textsf{O}[\vec{\CVar{c}}]}$
  \End
\end{codebox}

\begin{codebox}
  \Procname{$\proc{Make-Adjoint}(S)$} \li
    $\Stmt{\vec{\CVar{y}}, \vec{y} \coloneq \FVar{g}[\vec{\CVar{c}}, \vec{c}](\vec{x})} \gets S$\li
    \Return $\Stmt{\vec{\CVar{y}}', \vec{x} \coloneq \FVar{g}^\dagger[\vec{\CVar{c}}, \vec{c}](\vec{y})}$
  \End
\end{codebox}

\end{subfigure}%
\begin{subfigure}[t]{0.45\textwidth}

\begin{codebox}
  \Procname{$\proc{Synthesize-Adjoint}(\FVar{f})$} \li
    \textbf{assert} $\attrib{\FVar{f}}{effect} < m$\li
    $\FVar{f}^\dagger \gets \proc{New-Function}()$\li
    $\attrib{\FVar{f}^\dagger}{classical-in} \gets \attrib{\FVar{f}}{classical-in}$\li
    $\attrib{\FVar{f}^\dagger}{classical-out} \gets \attrib{\FVar{f}}{classical-out}$\li
    $\attrib{\FVar{f}^\dagger}{conserved} \gets \attrib{\FVar{f}}{conserved}$\li
    $\attrib{\FVar{f}^\dagger}{consumed} \gets \attrib{\FVar{f}}{produced}$\li
    $\attrib{\FVar{f}^\dagger}{produced} \gets \attrib{\FVar{f}}{consumed}$\li
    \For $S \in \attrib{\FVar{f}}{statements}$:\li
    \Do
      \If $\attrib{S}{classical-out} \neq \set{}$:\li
      \Then
        $S^\textsc{O} \gets \proc{Make-Classical}(S)$\li
        $\proc{Append}(\FVar{f}^\dagger, S^\textsc{O})$
      \End
    \End\li
    \For $S \in \id{reversed}(\attrib{\FVar{f}}{statements})$:\li
    \Do
      $S^\dag \gets \proc{Make-Adjoint}(S)$\li
      $\proc{Append}(\FVar{f}^\dagger, S^\dagger)$
    \End\li
    \Return $\FVar{f}^\dagger$
  \End
\end{codebox}
\end{subfigure}

%% file: impl-garbage.tex
\subsection{Uncomputation erasure}\label{sec:garbage}
Finally, we reintroduce the garbage monad.
This is done using \textit{garbage variables},
which are produced by calls in \textit{garbage mode} (denoted $\FVar{f}^\textsc{G}$)
and consumed by their adjoint (denoted $\FVar{f}^{\textsc{G}\dagger}$).

In a garbage mode function, the $\DisposeOp$ operation can be used
to throw a quantum variable into the garbage bin
or to propagate the garbage variable of another function.

Garbage-mode calls are key to our efficiency guarantees, as they avoid
the uncomputation and recomputation of intermediate variables within
the two components of a compute-uncompute pair. Instead, the compute
component computes such variables, and throws them into the garbage bin.
The uncompute component takes them back out of the garbage bin and
finally uncomputes them.

\subsubsection{Algorithm}
Our functions for garbage management are presented in
\cref{fig:garbage}.  In \cref{sec:uncomp}, we had already marked all
compute-uncompute pairs with $\proc{Connect-Garbage}$, but deferred
the explanations to the current section. $\proc{Connect-Garbage}$
introduces a new garbage variable, that is produced by a modified garbage-mode compute
statement and consumed by a modified garbage-mode uncompute statement.
The function $\id{op}^\textsc{G}$ is produced by calling $\proc{Erase-Uncomputation}(\id{op})$.

When $\proc{Synthesize-Adjoint}$ is performed on a garbage-mode call,
the garbage variable remains, but the roles of computation and
uncomputation are swapped. We also slightly adjust the
$\proc{Make-Adjoint}$ helper from \cref{fig:adjoint} such that it
preserves garbage variables.

Finally, we introduce $\proc{Erase-Uncomputation}(\FVar{f})$,
which is used to transform a pure function $\FVar{f}$
into the garbage-producing function $\FVar{f}^\textsc{G}$.
This transformation works by removing all instances of uncomputation from $\FVar{f}$
and replacing them with $\DisposeOp$. Uncomputation is identified as those statements that consume garbage.
It also transforms all other function calls in $\FVar{f}$ into garbage mode and disposes all garbage variables.
\begin{figure}\small
  \input{algo-garbage}
  \caption{Uncomputation erasure.}
  \label{fig:garbage}
\end{figure}
In the example below we have performed the transformation $\proc{Erase-Uncomputation}$ on $\FVar{maj}^\textsc{U}$:%
\begin{displaymath}
  \FVar{maj}^{\textsc{UG}}[\QVar{a}, \QVar{b}, \QVar{c}] \coloneq
  \left\{
  \begin{array}{rlrl}
    \QVar{t''}, g_0& \coloneq_p & \DupOp^\textsc{G}[\QVar{a}] & \\
                   & \coloneq_p & \DisposeOp(g_0) \\
    \QVar{t'}      & \coloneq_p & \DupOp[\QVar{t''}] & \\
    \QVar{x}, g_1  & \coloneq_p & \FVar{CX}^\textsc{G}[\QVar{b}](\QVar{t''}) & \\
                   & \coloneq_p & \DisposeOp(g_1) \\
    \QVar{r_0}, g_2 & \coloneq_p & \DupOp^\textsc{G}[\QVar{b}] & \opIf{\overline{\QVar{x}}} \\
                   & \coloneq_p & \DisposeOp(g_2) & \opIf{\overline{\QVar{x}}}\\
    \QVar{r_1}, g_3 & \coloneq_p & \DupOp^\textsc{G}[\QVar{c}] & \opIf{\QVar{x}} \\
                   & \coloneq_p & \DisposeOp(g_3) & \opIf{\QVar{x}} \\
    \QVar{r}, g_4  & \coloneq_p & \SelOp^\textsc{G}[\QVar{x}](\QVar{r_0}, \QVar{r_1}) & \\
                   & \coloneq_p & \DisposeOp(g_4) \\
                   & \coloneq_p & \DisposeOp(x) \\
    \QVar{r}, g_5  & \coloneq_p & \UnDupOp^\textsc{G}[\QVar{a}](\QVar{t'}) & \\
                   & \coloneq_p & \DisposeOp(g_5)
  \end{array}
  \right\}
  \triangleright \QVar{r}
\end{displaymath}
Note that $\FVar{maj}^\textsc{U}$ itself already contains garbage-mode
calls connecting the two compute-uncompute pairs for $\QVar{x}$ and
$\QVar{t}$, respectively. This is a detail we ignored in
\cref{sec:uncomp}. The compute-uncompute pair for $\QVar{t}$ we had
already simplified away earlier, so only the compute-uncompute pair
for $\QVar{x}$ remains. $\proc{Erase-Uncomputation}$ removes the
uncomputation of $\QVar{x}$ and replaces it with
$\DisposeOp{x}$ (lines 10-11). Because the uncomputation of $\QVar{x}$ produced the
variable $t'$, our algorithm has to now produce it in another way. It
achieves this by duplicating $\QVar{t''}$ before it is consumed (line 8).

All other statements from $\FVar{maj}^\textsc{U}$ are transformed into
garbage-mode calls and we propagate the new garbage variables by immediately
throwing them into the garbage bin using $\DisposeOp$.

There is still some opportunity for optimization. For example,
we can see that $\QVar{t'}$ is computed as $\DupOp[\QVar{t'']}$,
which in turn is $\DupOp[\QVar{a}]$,
and uncomputed as $\UnDupOp[\QVar{a}]$.
It is also unused, so it can be removed.
\begin{displaymath}
  \FVar{maj}^{\textsc{UG}}[\QVar{a}, \QVar{b}, \QVar{c}] \coloneq
  \left\{
  \begin{array}{rlrl}
    \QVar{t''}, g_0& \coloneq_p & \DupOp^\textsc{G}[\QVar{a}] & \\
                   & \coloneq_p & \DisposeOp(g_0) \\
    \QVar{x}, g_1  & \coloneq_p & \FVar{CX}^\textsc{G}[\QVar{b}](\QVar{t''}) & \\
                   & \coloneq_p & \DisposeOp(g_1) \\
    \QVar{r_0}, g_2 & \coloneq_p & \DupOp^\textsc{G}[\QVar{b}] & \opIf{\overline{\QVar{x}}} \\
                   & \coloneq_p & \DisposeOp(g_2) & \opIf{\overline{\QVar{x}}}\\
    \QVar{r_1}, g_3 & \coloneq_p & \DupOp^\textsc{G}[\QVar{c}] & \opIf{\QVar{x}} \\
                   & \coloneq_p & \DisposeOp(g_3) & \opIf{\QVar{x}} \\
    \QVar{r}, g_4  & \coloneq_p & \SelOp^\textsc{G}[\QVar{x}](\QVar{r_0}, \QVar{r_1}) & \\
                   & \coloneq_p & \DisposeOp(g_4) \\
                   & \coloneq_p & \DisposeOp(x)
  \end{array}
  \right\}
  \triangleright \QVar{r}
\end{displaymath}
We have shown how to connect computation-uncomputation pairs during uncomputation synthesis and modularly propagate their garbage. Next, we will argue that the resulting code is efficient.

\paragraph{Running time}
We will now show that with the use of garbage, for any source function $\FVar{f}$,
after uncomputation synthesis the asymptotic running time of $\FVar{f}^\textsc{U}$ remains the same as that of $\FVar{f}$,
even if we assume that every $\ForgetOp$ statement in $\FVar{f}$ takes \emph{constant} time to execute.

Consider a call from $\FVar{f}$ to $\FVar{callee}$. There are two cases:
\begin{itemize}
\item In $\proc{Synthesize-Uncomputation}$, the call to $\FVar{callee}$ was not undone.

This means that in the resulting function there is one call to $\FVar{callee}$,
and in $\FVar{f}^\dagger$ there is one call to $\FVar{callee}^\dagger$.
In both $\FVar{f}^\textsc{G}$ and $\FVar{f}^{\dagger\textsc{G}}$
there is one call to $\FVar{callee}^G$ instead.

\item In $\proc{Synthesize-Uncomputation}$, the call to $\FVar{callee}$ was undone.

This means that it was replaced with a call to $\FVar{callee}^\textsc{G}$
followed by a call to $\FVar{callee}^{\textsc{G}\dagger}$.

In $\FVar{f}^\dagger$ the two calls are swapped and adjoined,
meaning that there is one call to $\FVar{callee}^{\textsc{G}\dagger\dagger}=\FVar{callee}^\textsc{G}$
followed by a call to $\FVar{callee}^{\textsc{G}\dagger}$.

In both $\FVar{f}^\textsc{G}$ and $\FVar{f}^{\dagger\textsc{G}}$,
the call to $\FVar{callee}^{\textsc{G}\dagger}$ is removed
and only the call to $\FVar{callee}^\textsc{G}$ remains.
\end{itemize}

Therefore once we enter a function called in garbage mode, we never leave
garbage mode, and there is a 1-1 correspondence between function calls in the
original and the resulting program.

When we are in normal (not garbage) mode,
each call is either translated to a single call in normal mode,
or to two calls in garbage mode.

Thus the total number of calls performed during a call of $\FVar{f}^\textsc{U}$ after uncomputation synthesis
is at most double that of a call of the original function $\FVar{f}$.
The number of $\DupOp$ and $\UnDupOp$ statements inserted during the transformations
is at most a constant multiple of the number of arguments and results of function calls.
Therefore, the asymptotic running time indeed remains the same.

%% file: algo-garbage.tex
\begin{subfigure}[t]{0.43\textwidth}
\begin{codebox}
  \Procname{$\proc{Connect-Garbage}(C, U)$} \li
    $\id{op} \gets \attrib{C}{operation}$\li
    \textbf{assert} $\attrib{U}{operation} = \id{op}^\dagger$\li
    $\attrib{C}{operation} \gets \id{op}^\textsc{G}$\li
    $\attrib{U}{operation} \gets \id{op}^{\textsc{G}\dagger}$\li
    $\id{g} \gets \proc{Fresh-Garbage}()$\li
    $\attrib{C}{produced} \gets \attrib{C}{produced} \plusplus [g]$\li
    $\attrib{U}{consumed} \gets \attrib{U}{consumed} \plusplus [g]$
  \End
\end{codebox}

\begin{codebox}
  \Procname{$\proc{Propagate-Garbage}(S)$} \li
    \If $S$ does not produce garbage:\li
    \Then
      $\id{g} \gets \proc{Fresh-Garbage}()$\li
      $\attrib{S}{operation} \gets \attrib{S}{operation}^\textsf{G}$\li
      $\attrib{S}{produced} \gets \attrib{S}{produced} \plusplus [g]$\li
      $\proc{Insert-After}\left(S, \Stmt{\DisposeOp(g)}\right)$
    \End
  \End
\end{codebox}

\end{subfigure}%
\begin{subfigure}[t]{0.52\textwidth}

\begin{codebox}
  \Procname{$\proc{Erase-Uncomputation}(\FVar{f})$} \li
    \textbf{assert} $\attrib{\FVar{f}}{effect} < q$\li
    $\FVar{f}^\textsc{G} \gets \proc{Clone-Function}(\FVar{f})$\li
    \For $S \in \attrib{\FVar{f}^\textsc{G}}{statements}$:\li
    \Do
      \If $S \text{ produces garbage } g$:\li
      \Then
        $U \gets \id{consumer}(g)$\li
        \Comment{we remove $U$, so get its results from $S$}\li
        \For $(x, x') \in \id{zip}(\attrib{U}{produced}, \attrib{S}{consumed})$:\li
        \Do
          $\proc{Insert-Before}\left(S, \Stmt{x \coloneq \DupOp[x']}\right)$
        \End\li
        \For $x \in \attrib{U}{consumed}$:\li
        \Do
          $\proc{Insert-Before}\left(U, \Stmt{\DisposeOp(x)}\right)$
        \End\li
        $\proc{Remove}(U)$\li
        $\proc{Insert-After}\left(S, \Stmt{\DisposeOp(g)}\right)$
      \Else \If $\attrib{S}{operation} \neq \DisposeOp$:\li
        $\proc{Propagate-Garbage}(S)$
      \End
    \End\li
    \Return $\FVar{f}^\textsc{G}$
  \End
\end{codebox}
\end{subfigure}%

%% file: evaluation.tex
\section{Experimental evaluation}\label{sec:evaluation}
We now present an extensive experimental evaluation of our approach. Across a number of existing and new benchmarks, our evaluation aims to answer the following key research questions:
\begin{itemize}
\item[{\bf Q1}] \emph{Completeness}: Is our uncomputation synthesis approach effective when automatic uncomputation is intertwined with taking adjoints of subprograms?
\item[{\bf Q2}] \emph{Efficiency}: Is our modular uncomputation synthesis algorithm competitive with state-of-the-art (non-modular) synthesis approaches?
\end{itemize}
\paragraph{Choice of Baseline} We compare our approach to the state-of-the-art uncomputation synthesis algorithm Unqomp~\citep{unqomp}. Among other evaluation examples, we also include an evaluation on all of the original benchmark algorithms used by \citet{unqomp} to substantiate that Unqomp outperforms previous approaches. Overall, we are able to positively answer both questions above. We now proceed to discuss our evaluation.

\subsection{Implementation and experimental setup}
We implemented a compiler from a subset of the quantum programming language Silq~\citep{silq} to our quantum intermediate representation, in the process substantiating the utility of the IR as a target for high level quantum programming languages. We map automatic uncomputation in Silq to our $\ForgetOp$ operation, resulting in the first effective procedure for synthesizing explicit uncomputation code for Silq.

As Unqomp produces Qiskit circuits, we further implemented a simple lowering strategy for our IR, targeting Qiskit circuits via OpenQASM. In this way, our results can be directly compared to Unqomp results.

\paragraph{Unrolling} In contrast to our algorithm, Unqomp does not support recursion or variable-length quantum registers. Therefore, all evaluation examples we use to compare to the Unqomp baseline are similarly free of variable-length data, and we are able to completely inline and unroll our programs after modular uncomputation and adjoint synthesis are done.

\paragraph{Simplifications} After inlining and unrolling, we apply some standard simplifications, such as constant propagation, common subexpression elimination, and dead code elimination.

\paragraph{Register allocation} Because OpenQASM does not support dynamic qubit allocation, we further have to perform a register allocation pass. We use a simple strategy with a greedy heuristic that attempts to use the same registers for qubits that are distributed and selected in $\UnSelOp$ and $\SelOp$ operations.

Finally, we translate the resulting code directly into OpenQASM and import it as a Qiskit circuit. In all experiments, we first send the input qubits through a Hadamard transform to preclude constant folding. Like \citet{unqomp} does in the original Unqomp evaluation, we decompose all circuits (Unqomp and ours) into the universal gate set consisting of single-qubit unitary operations and two-qubit controlled-not (CX) gates.

\subsection{Q1: Completeness}
\input{table-unqomp-comparison-recursive}
We now demonstrate the benefits of our approach in cases where uncomputation and adjoint synthesis are intertwined. The benefit is illustrated on an example where the effect is pronounced. Consider the two example recursive programs in \cref{fig:forget-statement}. We implemented two recursive algorithms following the recursive structure of \proc{Iterate} and \proc{Etareti} respectively (discussed earlier), operating on quantum registers of size $10$. Instead of $y \gets x \oplus y$, the base case consists of an operation that is not its own adjoint, implemented using $CX$ gates. Single-qubit $X$ gates are used to initialize the variable $t$ to a list of $10$ qubits in state $\ket{1}$. We implemented versions of both algorithms in Silq (with native support for \kw{forget} statements) and Unqomp (implementing the \kw{forget} statement using Unqomp's automatic ancilla management).

Note that Unqomp does not have native support for inverting an ancilla circuit. This is analogous to the how the simulator distributed with the Silq reference implementation is not able to simulate the example with adjoints (at the time of writing), due to the way it takes the adjoint of functions containing \kw{forget} statements. Therefore, in the second benchmark, we invert Unqomp's ancilla circuits manually, in the canonical way, by first converting them to a Qiskit circuit (using Unqomp's uncomputation synthesis), inverting the Qiskit circuit, and transforming the result back into an ancilla circuit with the appropriate number of state and ancilla qubits.

For decomposition, we use Qiskit's preset pass manager with optimization level $0$ (no optimizations). \cref{table:unqomp-comparison-recursive} shows the results for this experiment. For the algorithm without adjoints, Unqomp and our approach both scale linearly in the recursion depth. In contrast, for the algorithm where we take an adjoint of every recursive call, the Unqomp implementation scales exponentially in the recursion depth, while our approach remains linear.

\subsection{Q2: Efficiency}
\input{table-unqomp-comparison}
We now demonstrate that our algorithms are competitive with the state of the art in terms of efficiency, namely the number of operations and qubits used, despite our IR allowing for more expressive quantum programs than circuits.

We implemented all benchmark programs from Unqomp in Silq. For Unqomp, we use the original implementations. Further, we implemented Shor's algorithm in both Silq and Unqomp\footnote{For Shor's algorithm, we had to patch Unqomp to add support for measurements and to fix a bug in the implementation of MCX.}. We have used the richer and more varied set of benchmark parameters from \citet{reqomp}, in both the {\bf small} and the {\bf big} variants. For Shor's algorithm, the {\bf small} variant factors $15$ using $4$ modulus bits, $2$ exponent bits, and a witness of $2$, while the {\bf big} variant factors $55$ using $6$ modulus bits, $12$ exponent bits, and a witness of $2$.

For decomposition, we use Qiskit's preset pass manager with optimization level $0$ (no optimizations). \cref{table:unqomp-comparison} shows the results for this experiment.

We also report results after applying further circuit-level optimizations in \cref{table:unqomp-comparison-optimized}. For those results, we use Qiskit's preset pass manager with the highest optimization level of $3$.

\input{table-unqomp-comparison-optimized}

\paragraph{Gate count} Our synthesis algorithms and simplifications outperform Unqomp on all benchmarks in terms of multi-qubit CX gates used. This is significant because multi-qubit gates are particularly hard to realize in hardware. Unqomp however often beats our results in terms of single-qubit gates.

We are ahead in overall gate count except for the polynomial Pauli rotation example with circuit optimizations enabled, where the Unqomp result appears to be more amenable to further circuit optimizations than what we produce.

\paragraph{Qubit count} Our qubit allocation strategy often matches Unqomp's ancilla allocation strategy and sometimes outperforms it. This is because with our more flexible approach, qubit allocation can benefit from previous simplifications such as constant folding. In all examples we consider, if Unqomp is better, it is only by at most one qubit.

Overall, these results demonstrate that our approach is competitive with state-of-the-art uncomputation algorithms, despite the fact that our algorithms work on a representation which allows for greater expressiveness of quantum computation than plain circuits do.

%% file: table-unqomp-comparison-recursive.tex
\begin{table}[t]
\caption{\label{table:unqomp-comparison-recursive}Recursive benchmarks. The best numbers are indicated in bold. In each row we report the number of single-qubit unitary gates (single), two-qubit controlled-not gates (CX), the total number of gates ($\text{Gates}=\text{single}+\text{CX}$) in a circuit produced using Unqomp or the new approach which is Ours.}
\begin{tabular}{l c rrrr c rrrr}
\toprule
&& \multicolumn{4}{c}{Unqomp} && \multicolumn{4}{c}{Ours}\\
Algorithm &&single&CX&Gates&Qubits&&single&CX&Gates&Qubits\\
\cmidrule(lr){3-6} \cmidrule(lr){8-11}
{\bf No adjoints}\\
$n=1$ && $\mathbf{40}$ & $96$ & $136$ & $\mathbf{21}$ && $46$ & $\mathbf{46}$ & $\mathbf{92}$ & $26$\\
$n=2$ && $\mathbf{60}$ & $160$ & $220$ & $\mathbf{31}$ && $66$ & $\mathbf{70}$ & $\mathbf{136}$ & $32$\\
$n=3$ && $\mathbf{80}$ & $224$ & $304$ & $41$ && $90$ & $\mathbf{94}$ & $\mathbf{184}$ & $\mathbf{38}$\\
$n=4$ && $\mathbf{100}$ & $288$ & $388$ & $51$ && $110$ & $\mathbf{118}$ & $\mathbf{228}$ & $\mathbf{44}$\\
$n=5$ && $\mathbf{120}$ & $352$ & $472$ & $61$ && $134$ & $\mathbf{142}$ & $\mathbf{276}$ & $\mathbf{50}$\\
$n=6$ && $\mathbf{140}$ & $416$ & $556$ & $71$ && $154$ & $\mathbf{166}$ & $\mathbf{320}$ & $\mathbf{56}$\\
$n=7$ && $\mathbf{160}$ & $480$ & $640$ & $81$ && $178$ & $\mathbf{190}$ & $\mathbf{368}$ & $\mathbf{62}$\\
$n=8$ && $\mathbf{180}$ & $544$ & $724$ & $91$ && $198$ & $\mathbf{214}$ & $\mathbf{412}$ & $\mathbf{68}$\\
$n=9$ && $\mathbf{200}$ & $608$ & $808$ & $101$ && $222$ & $\mathbf{238}$ & $\mathbf{460}$ & $\mathbf{74}$\\
$n=10$ && $\mathbf{220}$ & $672$ & $892$ & $111$ && $242$ & $\mathbf{262}$ & $\mathbf{504}$ & $\mathbf{80}$\\
{\bf Adjoints}\\
$n=1$ && $\mathbf{40}$ & $96$ & $136$ & $\mathbf{21}$ && $46$ & $\mathbf{46}$ & $\mathbf{92}$ & $26$\\
$n=2$ && $80$ & $224$ & $304$ & $\mathbf{31}$ && $\mathbf{68}$ & $\mathbf{70}$ & $\mathbf{138}$ & $32$\\
$n=3$ && $160$ & $480$ & $640$ & $41$ && $\mathbf{94}$ & $\mathbf{94}$ & $\mathbf{188}$ & $\mathbf{38}$\\
$n=4$ && $320$ & $992$ & $1312$ & $51$ && $\mathbf{116}$ & $\mathbf{118}$ & $\mathbf{234}$ & $\mathbf{44}$\\
$n=5$ && $640$ & $2016$ & $2656$ & $61$ && $\mathbf{142}$ & $\mathbf{142}$ & $\mathbf{284}$ & $\mathbf{50}$\\
$n=6$ && $1280$ & $4064$ & $5344$ & $71$ && $\mathbf{164}$ & $\mathbf{166}$ & $\mathbf{330}$ & $\mathbf{56}$\\
$n=7$ && $2560$ & $8160$ & $10720$ & $81$ && $\mathbf{190}$ & $\mathbf{190}$ & $\mathbf{380}$ & $\mathbf{62}$\\
$n=8$ && $5120$ & $16352$ & $21472$ & $91$ && $\mathbf{212}$ & $\mathbf{214}$ & $\mathbf{426}$ & $\mathbf{68}$\\
$n=9$ && $10240$ & $32736$ & $42976$ & $101$ && $\mathbf{238}$ & $\mathbf{238}$ & $\mathbf{476}$ & $\mathbf{74}$\\
$n=10$ && $20480$ & $65504$ & $85984$ & $111$ && $\mathbf{260}$ & $\mathbf{262}$ & $\mathbf{522}$ & $\mathbf{80}$\\
\bottomrule
\end{tabular}
\end{table}

%% file: table-unqomp-comparison.tex
\begin{table}[t]
\caption{\label{table:unqomp-comparison}Comparison to Unqomp without circuit-level optimizations. The best numbers are indicated in bold.}
\begin{tabular}{l c rrrr c rrrr}
\toprule
&& \multicolumn{4}{c}{Unqomp} && \multicolumn{4}{c}{Ours}\\
Algorithm &&single&CX&Gates&Qubits&&single&CX&Gates&Qubits\\
\cmidrule(lr){3-6} \cmidrule(lr){8-11}
{\bf small}\\
Adder && $\mathbf{288}$ & $400$ & $688$ & $36$ && $298$ & $\mathbf{155}$ & $\mathbf{453}$ & $\mathbf{35}$\\
Deutsch-Jozsa && $\mathbf{127}$ & $108$ & $235$ & $\mathbf{19}$ && $132$ & $\mathbf{58}$ & $\mathbf{190}$ & $20$\\
Grover's algorithm && $\mathbf{407}$ & $336$ & $743$ & $\mathbf{9}$ && $457$ & $\mathbf{179}$ & $\mathbf{636}$ & $10$\\
IntegerComparator && $184$ & $130$ & $314$ & $\mathbf{24}$ && $\mathbf{155}$ & $\mathbf{80}$ & $\mathbf{235}$ & $\mathbf{24}$\\
MCRY && $\mathbf{147}$ & $136$ & $283$ & $\mathbf{24}$ && $173$ & $\mathbf{68}$ & $\mathbf{241}$ & $\mathbf{24}$\\
MCX && $\mathbf{142}$ & $132$ & $274$ & $\mathbf{23}$ && $171$ & $\mathbf{67}$ & $\mathbf{238}$ & $24$\\
Multiplier && $675$ & $920$ & $1595$ & $\mathbf{24}$ && $\mathbf{407}$ & $\mathbf{220}$ & $\mathbf{627}$ & $\mathbf{24}$\\
PiecewiseLinearR && $1526$ & $1812$ & $3338$ & $13$ && $\mathbf{1281}$ & $\mathbf{743}$ & $\mathbf{2024}$ & $\mathbf{12}$\\
PolynomialPauliR && $\mathbf{607}$ & $660$ & $1267$ & $9$ && $642$ & $\mathbf{453}$ & $\mathbf{1095}$ & $\mathbf{6}$\\
WeightedAdder && $1499$ & $1618$ & $3117$ & $25$ && $\mathbf{754}$ & $\mathbf{403}$ & $\mathbf{1157}$ & $\mathbf{20}$\\
Shor's algorithm && $3610$ & $3760$ & $7370$ & $16$ && $\mathbf{83}$ & $\mathbf{29}$ & $\mathbf{112}$ & $\mathbf{8}$\\
{\bf big}\\
Adder && $\mathbf{2576}$ & $3568$ & $6144$ & $300$ && $2762$ & $\mathbf{1387}$ & $\mathbf{4149}$ & $\mathbf{299}$\\
Deutsch-Jozsa && $\mathbf{1387}$ & $1188$ & $2575$ & $\mathbf{199}$ && $1392$ & $\mathbf{598}$ & $\mathbf{1990}$ & $200$\\
Grover's algorithm && $\mathbf{6012}$ & $5100$ & $11112$ & $\mathbf{19}$ && $6564$ & $\mathbf{2603}$ & $\mathbf{9167}$ & $20$\\
IntegerComparator && $1746$ & $1192$ & $2938$ & $\mathbf{200}$ && $\mathbf{1381}$ & $\mathbf{690}$ & $\mathbf{2071}$ & $\mathbf{200}$\\
MCRY && $\mathbf{2591}$ & $2392$ & $4983$ & $\mathbf{400}$ && $2993$ & $\mathbf{1196}$ & $\mathbf{4189}$ & $\mathbf{400}$\\
MCX && $\mathbf{2586}$ & $2388$ & $4974$ & $\mathbf{399}$ && $2991$ & $\mathbf{1195}$ & $\mathbf{4186}$ & $400$\\
Multiplier && $7440$ & $10336$ & $17776$ & $\mathbf{79}$ && $\mathbf{4620}$ & $\mathbf{2431}$ & $\mathbf{7051}$ & $\mathbf{79}$\\
PiecewiseLinearR && $13858$ & $15792$ & $29650$ & $81$ && $\mathbf{11733}$ & $\mathbf{6739}$ & $\mathbf{18472}$ & $\mathbf{80}$\\
PolynomialPauliR && $\mathbf{51112}$ & $53144$ & $104256$ & $19$ && $51927$ & $\mathbf{45403}$ & $\mathbf{97330}$ & $\mathbf{11}$\\
WeightedAdder && $3542$ & $3794$ & $7336$ & $38$ && $\mathbf{2332}$ & $\mathbf{1248}$ & $\mathbf{3580}$ & $\mathbf{32}$\\
Shor's algorithm && $51462$ & $50692$ & $102154$ & $24$ && $\mathbf{22845}$ & $\mathbf{10673}$ & $\mathbf{33518}$ & $\mathbf{22}$\\
\bottomrule
\end{tabular}
\end{table}

%% file: table-unqomp-comparison-optimized.tex
\begin{table}[t]
\caption{\label{table:unqomp-comparison-optimized}Comparison to Unqomp after circuit-level optimizations. 
The best numbers are indicated in bold.}
\begin{tabular}{l c rrrr c rrrr}
\toprule
&& \multicolumn{4}{c}{Unqomp} && \multicolumn{4}{c}{Ours}\\
Algorithm &&single&CX&Gates&Qubits&&single&CX&Gates&Qubits\\
\cmidrule(lr){3-6} \cmidrule(lr){8-11}
{\bf small}\\
Adder && $156$ & $356$ & $512$ & $36$ && $\mathbf{152}$ & $\mathbf{149}$ & $\mathbf{301}$ & $\mathbf{35}$\\
Deutsch-Jozsa && $\mathbf{92}$ & $108$ & $200$ & $\mathbf{19}$ && $94$ & $\mathbf{55}$ & $\mathbf{149}$ & $20$\\
Grover's algorithm && $\mathbf{224}$ & $336$ & $560$ & $\mathbf{9}$ && $244$ & $\mathbf{155}$ & $\mathbf{399}$ & $10$\\
IntegerComparator && $\mathbf{101}$ & $130$ & $231$ & $\mathbf{24}$ && $\mathbf{101}$ & $\mathbf{77}$ & $\mathbf{178}$ & $\mathbf{24}$\\
MCRY && $\mathbf{102}$ & $136$ & $238$ & $\mathbf{24}$ && $128$ & $\mathbf{68}$ & $\mathbf{196}$ & $\mathbf{24}$\\
MCX && $\mathbf{99}$ & $132$ & $231$ & $\mathbf{23}$ && $127$ & $\mathbf{67}$ & $\mathbf{194}$ & $24$\\
Multiplier && $343$ & $840$ & $1183$ & $\mathbf{24}$ && $\mathbf{197}$ & $\mathbf{191}$ & $\mathbf{388}$ & $\mathbf{24}$\\
PiecewiseLinearR && $1102$ & $1784$ & $2886$ & $13$ && $\mathbf{799}$ & $\mathbf{682}$ & $\mathbf{1481}$ & $\mathbf{12}$\\
PolynomialPauliR && $\mathbf{290}$ & $556$ & $\mathbf{846}$ & $9$ && $434$ & $\mathbf{438}$ & $872$ & $\mathbf{6}$\\
WeightedAdder && $898$ & $1554$ & $2452$ & $25$ && $\mathbf{462}$ & $\mathbf{379}$ & $\mathbf{841}$ & $\mathbf{20}$\\
Shor's algorithm && $1985$ & $3720$ & $5705$ & $16$ && $\mathbf{51}$ & $\mathbf{29}$ & $\mathbf{80}$ & $\mathbf{8}$\\
{\bf big}\\
Adder && $1388$ & $3172$ & $4560$ & $300$ && $\mathbf{1384}$ & $\mathbf{1381}$ & $\mathbf{2765}$ & $\mathbf{299}$\\
Deutsch-Jozsa && $\mathbf{992}$ & $1188$ & $2180$ & $\mathbf{199}$ && $994$ & $\mathbf{595}$ & $\mathbf{1589}$ & $200$\\
Grover's algorithm && $\mathbf{3102}$ & $5100$ & $8202$ & $\mathbf{19}$ && $3472$ & $\mathbf{2208}$ & $\mathbf{5680}$ & $20$\\
IntegerComparator && $937$ & $1192$ & $2129$ & $\mathbf{200}$ && $\mathbf{893}$ & $\mathbf{687}$ & $\mathbf{1580}$ & $\mathbf{200}$\\
MCRY && $\mathbf{1794}$ & $2392$ & $4186$ & $\mathbf{400}$ && $2196$ & $\mathbf{1196}$ & $\mathbf{3392}$ & $\mathbf{400}$\\
MCX && $\mathbf{1791}$ & $2388$ & $4179$ & $\mathbf{399}$ && $2195$ & $\mathbf{1195}$ & $\mathbf{3390}$ & $400$\\
Multiplier && $3566$ & $9376$ & $12942$ & $\mathbf{79}$ && $\mathbf{1946}$ & $\mathbf{2182}$ & $\mathbf{4128}$ & $\mathbf{79}$\\
PiecewiseLinearR && $9276$ & $15440$ & $24716$ & $81$ && $\mathbf{7166}$ & $\mathbf{6397}$ & $\mathbf{13563}$ & $\mathbf{80}$\\
PolynomialPauliR && $\mathbf{25867}$ & $51680$ & $\mathbf{77547}$ & $19$ && $49123$ & $\mathbf{45362}$ & $94485$ & $\mathbf{11}$\\
WeightedAdder && $2072$ & $3602$ & $5674$ & $38$ && $\mathbf{1394}$ & $\mathbf{1187}$ & $\mathbf{2581}$ & $\mathbf{32}$\\
Shor's algorithm && $26078$ & $50482$ & $76560$ & $24$ && $\mathbf{11097}$ & $\mathbf{10522}$ & $\mathbf{21619}$ & $\mathbf{22}$\\
\bottomrule
\end{tabular}
\end{table}

%% file: conclusion.tex
\section{Conclusion}
We presented an intermediate representation (IR) which handles expressive quantum computation and provides uncomputation support together with efficient algorithms for modular uncomputation synthesis over that IR. The IR modularly captures the information needed for uncomputation and adjoint synthesis in a manner that allows our algorithms to naturally proceed on a function-by-function basis. Uncomputation synthesis uses the garbage monad to ensure that the resulting code does not perform redundant uncomputation. The resulting code can then be further simplified. 

We have built an end-to-end system consisting of a complete implementation of our IR and synthesis algorithms as well as translation from Silq to that IR and circuit generation from the IR. Our extensive evaluation shows that our system can efficiently handle benchmarks that are out of reach of existing methods for uncomputation synthesis, and that we are competitive with the state of the art at their own, more restricted, uncomputation benchmarks. These results indicate that it is in fact possible to benefit from the greater expressivity and safety offered by high-level languages without sacrificing efficiency.

%% file: acknowledgment.tex
\section*{Acknowledgment}
This research was partially funded by the Ministry of Education and Science of Bulgaria (support for INSAIT, part of the Bulgarian National Roadmap for Research Infrastructure).

%% file: proofs-uncomp.tex
First, we will show some invariants that guarantee that the resulting program is well-formed:

\begin{enumerate}
\item $\id{alive}$ contains the variables in scope at the point after the current statement.

This is maintained by line 14 of $\proc{Synthesize-Uncomputation}$.

\item All variables not in $\id{extended}$ are consumed exactly once, after being produced.

This invariant is maintained by the \For loop on lines 15-20 of $\proc{Undo-Statement}$.
In the case where a variable $x$ is not added to $\id{extended}$,
this is because it is not consumed before the current statement $S$,
and is instead consumed after it,
so we introduce a duplicate $x'$ that is used for the remainder of the function
(the parts that have already been visited).

\item $\proc{Ensure-Uncomputed}$ is only invoked on variables $x \not\in \id{alive}$.

We check the two call sites:
In $\proc{Synthesize-Uncomputation}$, $\attrib{S}{consumed} \cap \id{alive} = \set{}$,
as $S$ is considered to be before the current point.
In $\proc{Undo-Statement}$, $\attrib{C}{consumed} \cap \id{alive} = \set{}$,
as $C$ is before $S$.

\item After $\proc{Ensure-Uncomputed}$ is invoked on a variable $x$, we have $x \in \id{extended}$.

There are two cases we need to consider:
\begin{enumerate}
\item When $\id{producer}(x) \not\in \id{undone}$, $\proc{Undo-Statement}(S, \id{producer}(x))$ is invoked,
which adds $x$ to $\id{extended}$.
\item When $\id{producer}(x) \in \id{undone}$,
it is impossible for $\proc{Undo-Statement}(S, \id{producer}(x))$ to be in progress at this point
because we only ever go from a statement to a producer of one of its arguments.
Therefore $\proc{Undo-Statement}(S', \id{producer}(x))$ has already been performed
with some $S'$ at or after $S$, and as because variables never reenter scope,
$x$ was not in scope at the time, therefore it has added $x$ to $\id{extended}$.
\end{enumerate}

\item All variables in $\id{extended}$ are consumed once or twice:
\begin{itemize}
\item once after the current statement, by the inverse of a statement in $\id{undone}$
\item optionally once by or before the current statement
\end{itemize}

To prove this, we need to consider both updating $\id{extended}$ and visiting statements:
\begin{enumerate}
\item When a variable is added to $\id{extended}$ in $\proc{Undo-Statement}$,
it is not in scope,
meaning that it has been consumed before the current statement.
A second consumption has been added in the statement $U$.

\item When a statement $S$ that consumes $x \in \id{extended}$ is visited,
$S$ is either a $\ForgetOp$, in which case it is removed,
or it is not, in which case a $\DupOp$ of $x$ is created.
In either case, the result is that $x$ is no longer consumed at the point after $S$.
\end{enumerate}
\end{enumerate}

Now we can prove that the result is well-formed:
\begin{enumerate}
\item At the end of $\proc{Synthesize-Uncomputation}$, all variables are consumed exactly once.

This follows from the fact that at the end there are no statements before the current one,
so all variables in $\id{extended}$ are consumed once.

\item All previously-existing conserved uses remain valid.

This follows from the fact that we only ever extend lifetimes of variables.

\item All newly-introduced conserved uses are valid.

The only non-trivial cases are the uncomputation statement $U$
and the $\UnDupOp$ statements after it.
In both cases, we invoke $\proc{Ensure-Uncomputed}$ on all variables not in scope,
which ensures that the variables are added to $\id{extended}$.
They must be consumed twice: once before and once after $S$.
After the end of the transformation only the consumption after $S$ will remain,
so the conserved use is valid.
\end{enumerate}

What remains is to show that the uncomputation we synthesize is correct.

Suppose in $\proc{Undo-Statement}(S,C)$ we have a computation statement
\[
C = \big\{\QVar{y_1}, \ldots, \QVar{y_m} \coloneq \FVar{f}[\QVar{c_1}, \ldots, \QVar{c_k}](\QVar{x_1}, \ldots, \QVar{x_n})\big\}
\]

Then we insert a uncomputation statement
\[
U = \big\{\QVar{x'_1}, \ldots, \QVar{x'_n} \coloneq \FVar{f}^\dagger[\QVar{c_1}, \ldots, \QVar{c_k}](\QVar{y_1}, \ldots, \QVar{y_m})\big\}
\]

Here $\FVar{f}^\dagger$ is called with the same conserved arguments as $\FVar{f}$, and consumes the variables produced by $\FVar{f}$.
As $\FVar{f}$ is pure, and any use of $\QVar{c_i}$ and $\QVar{y_i}$ between computation and uncomputation necessarily conserves their value in the computational basis,
the resulting uncomputation is safe, i.e., $\QVar{x'_i}$ are equivalent to the original $\QVar{x_i}$.

We also add $\QVar{x_i}$ to $\id{extended}$, which results in $\DupOp$ statements being added before $C$.
Combined with the $\UnDupOp$ statements after $U$,
after uncomputation synthesis,
the complete result is non-consuming computation followed by its exact adjoint, which is non-producing uncomputation:
\begin{align*}
  & x''_1 \coloneq \DupOp[x_1] \quad\cdots\quad x''_n \coloneq \DupOp[x_n] \\
  & \QVar{y_1}, \ldots, \QVar{y_m} \coloneq \FVar{f}[\QVar{c_1}, \ldots, \QVar{c_k}](\QVar{x''_1}, \ldots, \QVar{x''_n}) \\
  & \ldots \\
  & \ldots \\
  & \QVar{x'_1}, \ldots, \QVar{x'_n} \coloneq \FVar{f}^\dagger[\QVar{c_1}, \ldots, \QVar{c_k}](\QVar{y_1}, \ldots, \QVar{y_m}) \\
  & \UnDupOp[x_1](x'_1) \quad\cdots\quad \UnDupOp[x_n](x'_n)
\end{align*}
Overall, our procedure synthesizes code with correctly scoped variables that still performs the same computation as the original code, but is free of the (hard-to-implement) $\ForgetOp$ statements.

%% file: main.bbl

\begin{thebibliography}{25}


\ifx \showCODEN    \undefined \def \showCODEN     #1{\unskip}     \fi
\ifx \showDOI      \undefined \def \showDOI       #1{#1}\fi
\ifx \showISBNx    \undefined \def \showISBNx     #1{\unskip}     \fi
\ifx \showISBNxiii \undefined \def \showISBNxiii  #1{\unskip}     \fi
\ifx \showISSN     \undefined \def \showISSN      #1{\unskip}     \fi
\ifx \showLCCN     \undefined \def \showLCCN      #1{\unskip}     \fi
\ifx \shownote     \undefined \def \shownote      #1{#1}          \fi
\ifx \showarticletitle \undefined \def \showarticletitle #1{#1}   \fi
\ifx \showURL      \undefined \def \showURL       {\relax}        \fi
\providecommand\bibfield[2]{#2}
\providecommand\bibinfo[2]{#2}
\providecommand\natexlab[1]{#1}
\providecommand\showeprint[2][]{arXiv:#2}

\bibitem[Aleksandrowicz et~al\mbox{.}(2019)]%
        {qiskit}
\bibfield{author}{\bibinfo{person}{Gadi Aleksandrowicz},
  \bibinfo{person}{Thomas Alexander}, \bibinfo{person}{Panagiotis Barkoutsos},
  \bibinfo{person}{Luciano Bello}, \bibinfo{person}{Yael Ben-Haim},
  \bibinfo{person}{David Bucher}, \bibinfo{person}{Francisco~Jose
  Cabrera-Hernández}, \bibinfo{person}{Jorge Carballo-Franquis},
  \bibinfo{person}{Adrian Chen}, \bibinfo{person}{Chun-Fu Chen},
  \bibinfo{person}{Jerry~M. Chow}, \bibinfo{person}{Antonio~D.
  Córcoles-Gonzales}, \bibinfo{person}{Abigail~J. Cross},
  \bibinfo{person}{Andrew Cross}, \bibinfo{person}{Juan Cruz-Benito},
  \bibinfo{person}{Chris Culver}, \bibinfo{person}{Salvador De La~Puente
  González}, \bibinfo{person}{Enrique De~La Torre}, \bibinfo{person}{Delton
  Ding}, \bibinfo{person}{Eugene Dumitrescu}, \bibinfo{person}{Ivan Duran},
  \bibinfo{person}{Pieter Eendebak}, \bibinfo{person}{Mark Everitt},
  \bibinfo{person}{Ismael~Faro Sertage}, \bibinfo{person}{Albert Frisch},
  \bibinfo{person}{Andreas Fuhrer}, \bibinfo{person}{Jay Gambetta},
  \bibinfo{person}{Borja~Godoy Gago}, \bibinfo{person}{Juan Gomez-Mosquera},
  \bibinfo{person}{Donny Greenberg}, \bibinfo{person}{Ikko Hamamura},
  \bibinfo{person}{Vojtech Havlicek}, \bibinfo{person}{Joe Hellmers},
  \bibinfo{person}{Łukasz Herok}, \bibinfo{person}{Hiroshi Horii},
  \bibinfo{person}{Shaohan Hu}, \bibinfo{person}{Takashi Imamichi},
  \bibinfo{person}{Toshinari Itoko}, \bibinfo{person}{Ali Javadi-Abhari},
  \bibinfo{person}{Naoki Kanazawa}, \bibinfo{person}{Anton Karazeev},
  \bibinfo{person}{Kevin Krsulich}, \bibinfo{person}{Peng Liu},
  \bibinfo{person}{Yang Luh}, \bibinfo{person}{Yunho Maeng},
  \bibinfo{person}{Manoel Marques}, \bibinfo{person}{Francisco~Jose
  Martín-Fernández}, \bibinfo{person}{Douglas~T. McClure},
  \bibinfo{person}{David McKay}, \bibinfo{person}{Srujan Meesala},
  \bibinfo{person}{Antonio Mezzacapo}, \bibinfo{person}{Nikolaj Moll},
  \bibinfo{person}{Diego~Moreda Rodríguez}, \bibinfo{person}{Giacomo
  Nannicini}, \bibinfo{person}{Paul Nation}, \bibinfo{person}{Pauline
  Ollitrault}, \bibinfo{person}{Lee~James O'Riordan}, \bibinfo{person}{Hanhee
  Paik}, \bibinfo{person}{Jesús Pérez}, \bibinfo{person}{Anna Phan},
  \bibinfo{person}{Marco Pistoia}, \bibinfo{person}{Viktor Prutyanov},
  \bibinfo{person}{Max Reuter}, \bibinfo{person}{Julia Rice},
  \bibinfo{person}{Abdón~Rodríguez Davila}, \bibinfo{person}{Raymond
  Harry~Putra Rudy}, \bibinfo{person}{Mingi Ryu}, \bibinfo{person}{Ninad
  Sathaye}, \bibinfo{person}{Chris Schnabel}, \bibinfo{person}{Eddie Schoute},
  \bibinfo{person}{Kanav Setia}, \bibinfo{person}{Yunong Shi},
  \bibinfo{person}{Adenilton Silva}, \bibinfo{person}{Yukio Siraichi},
  \bibinfo{person}{Seyon Sivarajah}, \bibinfo{person}{John~A. Smolin},
  \bibinfo{person}{Mathias Soeken}, \bibinfo{person}{Hitomi Takahashi},
  \bibinfo{person}{Ivano Tavernelli}, \bibinfo{person}{Charles Taylor},
  \bibinfo{person}{Pete Taylour}, \bibinfo{person}{Kenso Trabing},
  \bibinfo{person}{Matthew Treinish}, \bibinfo{person}{Wes Turner},
  \bibinfo{person}{Desiree Vogt-Lee}, \bibinfo{person}{Christophe Vuillot},
  \bibinfo{person}{Jonathan~A. Wildstrom}, \bibinfo{person}{Jessica Wilson},
  \bibinfo{person}{Erick Winston}, \bibinfo{person}{Christopher Wood},
  \bibinfo{person}{Stephen Wood}, \bibinfo{person}{Stefan Wörner},
  \bibinfo{person}{Ismail~Yunus Akhalwaya}, {and} \bibinfo{person}{Christa
  Zoufal}.} \bibinfo{year}{2019}\natexlab{}.
\newblock \bibinfo{booktitle}{\emph{{Qiskit: An Open-source Framework for
  Quantum Computing}}}.
\newblock
\urldef\tempurl%
\url{https://doi.org/10.5281/zenodo.2562111}
\showDOI{\tempurl}


\bibitem[Amy et~al\mbox{.}(2017)]%
        {amy2017verified}
\bibfield{author}{\bibinfo{person}{Matthew Amy}, \bibinfo{person}{Martin
  Roetteler}, {and} \bibinfo{person}{Krysta~M Svore}.}
  \bibinfo{year}{2017}\natexlab{}.
\newblock \showarticletitle{Verified compilation of space-efficient reversible
  circuits}. In \bibinfo{booktitle}{\emph{International Conference on Computer
  Aided Verification}}. Springer, \bibinfo{pages}{3--21}.
\newblock


\bibitem[Bennett(1973)]%
        {bennett_logical_1973}
\bibfield{author}{\bibinfo{person}{C.~H. Bennett}.}
  \bibinfo{year}{1973}\natexlab{}.
\newblock \showarticletitle{Logical Reversibility of Computation}.
\newblock \bibinfo{journal}{\emph{IBM Journal of Research and Development}}
  \bibinfo{volume}{17}, \bibinfo{number}{6} (\bibinfo{date}{Nov.}
  \bibinfo{year}{1973}), \bibinfo{pages}{525--532}.
\newblock
\showISSN{0018-8646}
\urldef\tempurl%
\url{https://doi.org/10.1147/rd.176.0525}
\showDOI{\tempurl}


\bibitem[Bennett(1989)]%
        {bennett_timespace_1989}
\bibfield{author}{\bibinfo{person}{Charles~H. Bennett}.}
  \bibinfo{year}{1989}\natexlab{}.
\newblock \showarticletitle{Time/{Space} {Trade}-{Offs} for {Reversible}
  {Computation}}.
\newblock \bibinfo{journal}{\emph{SIAM J. Comput.}} \bibinfo{volume}{18},
  \bibinfo{number}{4} (\bibinfo{date}{Aug.} \bibinfo{year}{1989}),
  \bibinfo{pages}{766--776}.
\newblock
\showISSN{0097-5397}
\urldef\tempurl%
\url{https://doi.org/10.1137/0218053}
\showDOI{\tempurl}
\newblock
\shownote{Publisher: Society for Industrial and Applied Mathematics}.


\bibitem[Bennett et~al\mbox{.}(1993)]%
        {quantum_teleportation}
\bibfield{author}{\bibinfo{person}{Charles~H. Bennett}, \bibinfo{person}{Gilles
  Brassard}, \bibinfo{person}{Claude Cr\'epeau}, \bibinfo{person}{Richard
  Jozsa}, \bibinfo{person}{Asher Peres}, {and} \bibinfo{person}{William~K.
  Wootters}.} \bibinfo{year}{1993}\natexlab{}.
\newblock \showarticletitle{Teleporting an unknown quantum state via dual
  classical and Einstein-Podolsky-Rosen channels}.
\newblock \bibinfo{journal}{\emph{Phys. Rev. Lett.}}  \bibinfo{volume}{70}
  (\bibinfo{date}{Mar} \bibinfo{year}{1993}), \bibinfo{pages}{1895--1899}.
\newblock
Issue 13.
\urldef\tempurl%
\url{https://doi.org/10.1103/PhysRevLett.70.1895}
\showDOI{\tempurl}


\bibitem[Bichsel et~al\mbox{.}(2020)]%
        {silq}
\bibfield{author}{\bibinfo{person}{Benjamin Bichsel},
  \bibinfo{person}{Maximilian Baader}, \bibinfo{person}{Timon Gehr}, {and}
  \bibinfo{person}{Martin Vechev}.} \bibinfo{year}{2020}\natexlab{}.
\newblock \showarticletitle{Silq: A High-Level Quantum Language with Safe
  Uncomputation and Intuitive Semantics}. In
  \bibinfo{booktitle}{\emph{Proceedings of the 41st ACM SIGPLAN Conference on
  Programming Language Design and Implementation}} (London, UK)
  \emph{(\bibinfo{series}{PLDI 2020})}. \bibinfo{publisher}{Association for
  Computing Machinery}, \bibinfo{address}{New York, NY, USA},
  \bibinfo{pages}{286–300}.
\newblock
\showISBNx{9781450376136}
\urldef\tempurl%
\url{https://doi.org/10.1145/3385412.3386007}
\showDOI{\tempurl}


\bibitem[Fu et~al\mbox{.}(2020)]%
        {protoquipperd}
\bibfield{author}{\bibinfo{person}{Peng Fu}, \bibinfo{person}{Kohei Kishida},
  \bibinfo{person}{Neil~J Ross}, {and} \bibinfo{person}{Peter Selinger}.}
  \bibinfo{year}{2020}\natexlab{}.
\newblock \showarticletitle{A tutorial introduction to quantum circuit
  programming in dependently typed Proto-Quipper}. In
  \bibinfo{booktitle}{\emph{Reversible Computation: 12th International
  Conference, RC 2020, Oslo, Norway, July 9-10, 2020, Proceedings 12}}.
  Springer, \bibinfo{pages}{153--168}.
\newblock


\bibitem[Green et~al\mbox{.}(2013)]%
        {quipper}
\bibfield{author}{\bibinfo{person}{Alexander~S. Green},
  \bibinfo{person}{Peter~LeFanu Lumsdaine}, \bibinfo{person}{Neil~J. Ross},
  \bibinfo{person}{Peter Selinger}, {and} \bibinfo{person}{Benoît Valiron}.}
  \bibinfo{year}{2013}\natexlab{}.
\newblock \showarticletitle{Quipper: a scalable quantum programming language}.
  In \bibinfo{booktitle}{\emph{PLDI'13}}. \bibinfo{publisher}{ACM Press},
  \bibinfo{address}{Seattle, Washington, USA}.
\newblock
\showISBNx{978-1-4503-2014-6}
\urldef\tempurl%
\url{https://doi.org/10.1145/2491956.2462177}
\showDOI{\tempurl}


\bibitem[Heunen and Kaarsgaard(2022)]%
        {heunen_quantum_2022}
\bibfield{author}{\bibinfo{person}{Chris Heunen} {and} \bibinfo{person}{Robin
  Kaarsgaard}.} \bibinfo{year}{2022}\natexlab{}.
\newblock \showarticletitle{Quantum information effects}.
\newblock \bibinfo{journal}{\emph{Proceedings of the ACM on Programming
  Languages}} \bibinfo{volume}{6}, \bibinfo{number}{POPL} (\bibinfo{date}{Jan.}
  \bibinfo{year}{2022}), \bibinfo{pages}{2:1--2:27}.
\newblock
\urldef\tempurl%
\url{https://doi.org/10.1145/3498663}
\showDOI{\tempurl}


\bibitem[James and Sabry(2012)]%
        {james_information_2012}
\bibfield{author}{\bibinfo{person}{Roshan~P. James} {and} \bibinfo{person}{Amr
  Sabry}.} \bibinfo{year}{2012}\natexlab{}.
\newblock \showarticletitle{Information effects}. In
  \bibinfo{booktitle}{\emph{Proceedings of the 39th annual {ACM}
  {SIGPLAN}-{SIGACT} symposium on {Principles} of programming languages}}
  \emph{(\bibinfo{series}{{POPL} '12})}. \bibinfo{publisher}{Association for
  Computing Machinery}, \bibinfo{address}{New York, NY, USA},
  \bibinfo{pages}{73--84}.
\newblock
\showISBNx{978-1-4503-1083-3}
\urldef\tempurl%
\url{https://doi.org/10.1145/2103656.2103667}
\showDOI{\tempurl}


\bibitem[Kumar~Pati and Braunstein(2000)]%
        {kumar_pati_impossibility_2000}
\bibfield{author}{\bibinfo{person}{Arun Kumar~Pati} {and}
  \bibinfo{person}{Samuel~L. Braunstein}.} \bibinfo{year}{2000}\natexlab{}.
\newblock \showarticletitle{Impossibility of deleting an unknown quantum
  state}.
\newblock \bibinfo{journal}{\emph{Nature}} \bibinfo{volume}{404},
  \bibinfo{number}{6774} (\bibinfo{date}{March} \bibinfo{year}{2000}),
  \bibinfo{pages}{164--165}.
\newblock
\showISSN{1476-4687}
\urldef\tempurl%
\url{https://doi.org/10.1038/404130b0}
\showDOI{\tempurl}
\newblock
\shownote{Publisher: Nature Publishing Group}.


\bibitem[Lecerf(1963)]%
        {lecerf_machines_1963}
\bibfield{author}{\bibinfo{person}{Yves Lecerf}.}
  \bibinfo{year}{1963}\natexlab{}.
\newblock \showarticletitle{Machines de {Turing} r\'{e}versibles.
  {R\'{e}cursive} insolubilit\'{e} en $n \in \mathbb{N}$ de l`\'{e}quation $u =
  \theta^n u$, o\`{u} $\theta$ est un isomorphisme de codes}.
\newblock \bibinfo{journal}{\emph{Comptes Rendus Hebdomadaires des Seances de
  L'academie des Sciences}}  \bibinfo{volume}{257} (\bibinfo{year}{1963}),
  \bibinfo{pages}{2597--2600}.
\newblock


\bibitem[Lubinski et~al\mbox{.}(2022)]%
        {qir}
\bibfield{author}{\bibinfo{person}{Thomas Lubinski}, \bibinfo{person}{Cassandra
  Granade}, \bibinfo{person}{Amos Anderson}, \bibinfo{person}{Alan Geller},
  \bibinfo{person}{Martin Roetteler}, \bibinfo{person}{Andrei Petrenko}, {and}
  \bibinfo{person}{Bettina Heim}.} \bibinfo{year}{2022}\natexlab{}.
\newblock \showarticletitle{Advancing hybrid quantum–classical computation
  with real-time execution}.
\newblock \bibinfo{journal}{\emph{Frontiers in Physics}}  \bibinfo{volume}{10}
  (\bibinfo{year}{2022}).
\newblock
\showISSN{2296-424X}
\urldef\tempurl%
\url{https://doi.org/10.3389/fphy.2022.940293}
\showDOI{\tempurl}


\bibitem[Microsoft(2020)]%
        {qsharp}
\bibfield{author}{\bibinfo{person}{Microsoft}.}
  \bibinfo{year}{2020}\natexlab{}.
\newblock \bibinfo{booktitle}{\emph{Q\# Language Specification}}.
\newblock
\urldef\tempurl%
\url{https://github.com/microsoft/qsharp-language/tree/main/Specifications/Language#q-language}
\showURL{%
Retrieved Jul 7, 2023 from \tempurl}


\bibitem[Nielsen and Chuang(2000)]%
        {nielsenchuang}
\bibfield{author}{\bibinfo{person}{Michael~A. Nielsen} {and}
  \bibinfo{person}{Isaac~L. Chuang}.} \bibinfo{year}{2000}\natexlab{}.
\newblock \bibinfo{booktitle}{\emph{Quantum Computation and Quantum
  Information}}.
\newblock \bibinfo{publisher}{Cambridge University Press}.
\newblock


\bibitem[Paradis et~al\mbox{.}(2021)]%
        {unqomp}
\bibfield{author}{\bibinfo{person}{Anouk Paradis}, \bibinfo{person}{Benjamin
  Bichsel}, \bibinfo{person}{Samuel Steffen}, {and} \bibinfo{person}{Martin
  Vechev}.} \bibinfo{year}{2021}\natexlab{}.
\newblock \showarticletitle{Unqomp: Synthesizing Uncomputation in Quantum
  Circuits}. In \bibinfo{booktitle}{\emph{Proceedings of the 42nd ACM SIGPLAN
  International Conference on Programming Language Design and Implementation}}
  (Virtual, Canada) \emph{(\bibinfo{series}{PLDI 2021})}.
  \bibinfo{publisher}{Association for Computing Machinery},
  \bibinfo{address}{New York, NY, USA}, \bibinfo{pages}{222–236}.
\newblock
\showISBNx{9781450383912}
\urldef\tempurl%
\url{https://doi.org/10.1145/3453483.3454040}
\showDOI{\tempurl}


\bibitem[Paradis et~al\mbox{.}(2022)]%
        {reqomp}
\bibfield{author}{\bibinfo{person}{Anouk Paradis}, \bibinfo{person}{Benjamin
  Bichsel}, {and} \bibinfo{person}{Martin Vechev}.}
  \bibinfo{year}{2022}\natexlab{}.
\newblock \bibinfo{title}{Reqomp: Space-constrained Uncomputation for Quantum
  Circuits}.
\newblock
\newblock
\showeprint[arxiv]{2212.10395}~[quant-ph]


\bibitem[Parent et~al\mbox{.}(2015)]%
        {parent2015reversible}
\bibfield{author}{\bibinfo{person}{Alex Parent}, \bibinfo{person}{Martin
  Roetteler}, {and} \bibinfo{person}{Krysta~M. Svore}.}
  \bibinfo{year}{2015}\natexlab{}.
\newblock \showarticletitle{Reversible circuit compilation with space
  constraints}.
\newblock \bibinfo{journal}{\emph{CoRR}}  \bibinfo{volume}{abs/1510.00377}
  (\bibinfo{year}{2015}).
\newblock
\showeprint[arXiv]{1510.00377}
\urldef\tempurl%
\url{http://arxiv.org/abs/1510.00377}
\showURL{%
\tempurl}


\bibitem[Paykin et~al\mbox{.}(2017)]%
        {qwire}
\bibfield{author}{\bibinfo{person}{Jennifer Paykin}, \bibinfo{person}{Robert
  Rand}, {and} \bibinfo{person}{Steve Zdancewic}.}
  \bibinfo{year}{2017}\natexlab{}.
\newblock \showarticletitle{{QWIRE}: a core language for quantum circuits}. In
  \bibinfo{booktitle}{\emph{POPL'17}}. \bibinfo{publisher}{ACM Press},
  \bibinfo{address}{Paris, France}.
\newblock
\showISBNx{978-1-4503-4660-3}
\urldef\tempurl%
\url{https://doi.org/10.1145/3009837.3009894}
\showDOI{\tempurl}


\bibitem[Peduri et~al\mbox{.}(2022)]%
        {qssa}
\bibfield{author}{\bibinfo{person}{Anurudh Peduri}, \bibinfo{person}{Siddharth
  Bhat}, {and} \bibinfo{person}{Tobias Grosser}.}
  \bibinfo{year}{2022}\natexlab{}.
\newblock \showarticletitle{QSSA: An SSA-Based IR for Quantum Computing}. In
  \bibinfo{booktitle}{\emph{Proceedings of the 31st ACM SIGPLAN International
  Conference on Compiler Construction}} (Seoul, South Korea)
  \emph{(\bibinfo{series}{CC 2022})}. \bibinfo{publisher}{Association for
  Computing Machinery}, \bibinfo{address}{New York, NY, USA},
  \bibinfo{pages}{2–14}.
\newblock
\showISBNx{9781450391832}
\urldef\tempurl%
\url{https://doi.org/10.1145/3497776.3517772}
\showDOI{\tempurl}


\bibitem[Rand et~al\mbox{.}(2019)]%
        {reqwire}
\bibfield{author}{\bibinfo{person}{Robert Rand}, \bibinfo{person}{Jennifer
  Paykin}, \bibinfo{person}{Dong-Ho Lee}, {and} \bibinfo{person}{Steve
  Zdancewic}.} \bibinfo{year}{2019}\natexlab{}.
\newblock \showarticletitle{{ReQWIRE}: {Reasoning} about {Reversible} {Quantum}
  {Circuits}}.
\newblock \bibinfo{journal}{\emph{Electronic Proceedings in Theoretical
  Computer Science}}  \bibinfo{volume}{287} (\bibinfo{date}{Jan.}
  \bibinfo{year}{2019}), \bibinfo{pages}{299--312}.
\newblock
\showISSN{2075-2180}
\urldef\tempurl%
\url{https://doi.org/10.4204/EPTCS.287.17}
\showDOI{\tempurl}
\newblock
\shownote{arXiv: 1901.10118}.


\bibitem[Shor(1997)]%
        {shor}
\bibfield{author}{\bibinfo{person}{Peter~W. Shor}.}
  \bibinfo{year}{1997}\natexlab{}.
\newblock \showarticletitle{Polynomial-Time Algorithms for Prime Factorization
  and Discrete Logarithms on a Quantum Computer}.
\newblock \bibinfo{journal}{\emph{SIAM J. Comput.}} \bibinfo{volume}{26},
  \bibinfo{number}{5} (\bibinfo{date}{Oct.} \bibinfo{year}{1997}),
  \bibinfo{pages}{1484–1509}.
\newblock
\showISSN{1095-7111}
\urldef\tempurl%
\url{https://doi.org/10.1137/s0097539795293172}
\showDOI{\tempurl}


\bibitem[Toffoli(1980)]%
        {toffoli_reversible_1980}
\bibfield{author}{\bibinfo{person}{Tommaso Toffoli}.}
  \bibinfo{year}{1980}\natexlab{}.
\newblock \showarticletitle{Reversible computing}. In
  \bibinfo{booktitle}{\emph{Automata, {Languages} and {Programming}}}
  \emph{(\bibinfo{series}{Lecture Notes in Computer Science})},
  \bibfield{editor}{\bibinfo{person}{Jaco de~Bakker} {and} \bibinfo{person}{Jan
  van Leeuwen}} (Eds.). \bibinfo{publisher}{Springer},
  \bibinfo{address}{Berlin, Heidelberg}, \bibinfo{pages}{632--644}.
\newblock
\showISBNx{978-3-540-39346-7}
\urldef\tempurl%
\url{https://doi.org/10.1007/3-540-10003-2_104}
\showDOI{\tempurl}


\bibitem[Voichick et~al\mbox{.}(2023)]%
        {qunity}
\bibfield{author}{\bibinfo{person}{Finn Voichick}, \bibinfo{person}{Liyi Li},
  \bibinfo{person}{Robert Rand}, {and} \bibinfo{person}{Michael Hicks}.}
  \bibinfo{year}{2023}\natexlab{}.
\newblock \showarticletitle{Qunity: A Unified Language for Quantum and
  Classical Computing}.
\newblock \bibinfo{journal}{\emph{Proc. ACM Program. Lang.}}
  \bibinfo{volume}{7}, \bibinfo{number}{POPL}, Article \bibinfo{articleno}{32}
  (\bibinfo{date}{jan} \bibinfo{year}{2023}), \bibinfo{numpages}{31}~pages.
\newblock
\urldef\tempurl%
\url{https://doi.org/10.1145/3571225}
\showDOI{\tempurl}


\bibitem[Yuan and Carbin(2022)]%
        {tower}
\bibfield{author}{\bibinfo{person}{Charles Yuan} {and} \bibinfo{person}{Michael
  Carbin}.} \bibinfo{year}{2022}\natexlab{}.
\newblock \showarticletitle{Tower: Data Structures in Quantum Superposition}.
\newblock \bibinfo{journal}{\emph{Proc. ACM Program. Lang.}}
  \bibinfo{volume}{6}, \bibinfo{number}{OOPSLA2}, Article
  \bibinfo{articleno}{134} (\bibinfo{date}{oct} \bibinfo{year}{2022}),
  \bibinfo{numpages}{30}~pages.
\newblock
\urldef\tempurl%
\url{https://doi.org/10.1145/3563297}
\showDOI{\tempurl}


\end{thebibliography}
